\documentclass[sigconf,screen]{acmart}

\usepackage{amsmath,amsfonts}
\usepackage{algorithmic}
\usepackage{graphicx,calc}
\usepackage{textcomp}
\usepackage{xcolor}
\usepackage{listings}
\usepackage[utf8]{inputenc}
\usepackage[flushleft]{threeparttable}
\usepackage{caption}
\usepackage{subcaption}
\usepackage{balance}
\usepackage{wrapfig}
\usepackage{textcomp}
\usepackage{mathtools,amsthm}
\usepackage{tcolorbox}
\usepackage{xstring}
\usepackage{comment}
\usepackage{multicol}
\usepackage{enumitem}

\graphicspath{{images/}}

\newcommand{\squishlist}{
  \begin{itemize}[noitemsep, nolistsep, leftmargin=*]
  \setlength{\itemsep}{-0pt}
}
\newcommand{\squishstart}{
  \begin{itemize}
}
\newcommand{\squishend}{
  \end{itemize}
}
\newcommand{\commentt}[1]{}
\newcommand{\bettersim}{{\raise.17ex\hbox{$\scriptstyle\sim$}}}

\setcopyright{rightsretained}
\acmPrice{}
\acmDOI{10.1145/3533767.3534401}
\acmYear{2022}
\copyrightyear{2022}
\acmSubmissionID{issta22main-p382-p}
\acmISBN{978-1-4503-9379-9/22/07}
\acmConference[ISSTA '22]{Proceedings of the 31st ACM SIGSOFT International Symposium on Software Testing and Analysis}{July 18--22, 2022}{Virtual, South Korea}
\acmBooktitle{Proceedings of the 31st ACM SIGSOFT International Symposium on Software Testing and Analysis (ISSTA '22), July 18--22, 2022, Virtual, South Korea}



\begin{document}

\lstset{
  basicstyle=\fontsize{6}{6}\ttfamily\bfseries,
  numbers=left,
  numbersep=3pt,
  numberstyle=\tiny,
  xleftmargin=0.1in,
}
\newcommand{\alex}[1]{{\color{red}\bf [AO: #1]}}
\newcommand{\myeongsoo}[1]{{\color{orange}\bf [MK: #1]}}
\newcommand{\qi}[1]{{\color{blue}\bf [QX: #1]}}
\newcommand{\qifirst}[1]{{\color{red}\bf [QX: #1]}}
\newcommand{\saurabh}[1]{{\color{cyan}\bf [SS: #1]}}

\title{Automated Test Generation for REST APIs: No Time to Rest Yet}

\author{Myeongsoo Kim}
\affiliation{
  \institution{Georgia Institute of Technology}
  \country{USA}
}
\email{mkim754@gatech.edu}

\author{Qi Xin}
\affiliation{
  \institution{Wuhan University}
  \country{China}
}
\email{qxin@whu.edu.cn}

\author{Saurabh Sinha}
\affiliation{
  \institution{IBM T.J. Watson Research Center}
  \country{USA}
}
\email{sinhas@us.ibm.com}

\author{Alessandro Orso}
\affiliation{
  \institution{Georgia Institute of Technology}
  \country{USA}
}
\email{orso@cc.gatech.edu}

\begin{abstract}
Modern web services routinely provide REST APIs for clients to access their functionality. 
These APIs present unique challenges and opportunities for automated testing, driving the recent development of many techniques and tools that generate test cases for API endpoints using various strategies.
Understanding how these techniques compare to one another is difficult, as they have been evaluated on different benchmarks and using different metrics.
To fill this gap, we performed an empirical study aimed to understand the landscape in automated testing of REST APIs and guide future research in this area.
We first identified, through a systematic selection process, a set of 10 state-of-the-art REST API testing tools that included tools developed by both researchers and practitioners.
We then applied these tools to a benchmark of 20 real-world open-source RESTful services and analyzed their performance in terms of code coverage achieved and unique failures triggered.
This analysis allowed us to identify strengths, weaknesses, and limitations of the tools considered and of their underlying strategies, as well as implications of our findings for future research in this area.
\end{abstract}


\maketitle

\section{Introduction}
\label{sec:introduction}

The last decade has seen a tremendous growth in the availability of web APIs---APIs that provide access to a service through a web interface.
This increased popularity has been driven by various industry trends, including the advent of cloud computing, the broad adoption of microservices~\cite{newman:2015}, and newer value propositions enabled by the ``API economy''.
A majority of modern web APIs adhere to the REpresentational State Transfer (REST) architectural style~\cite{fielding2000architectural} and are referred to as RESTful APIs, whose popularity is reflected in the availability of thousands of APIs in public directories (e.g., ProgrammableWeb~\cite{progweb} and APIs guru~\cite{apisguru}).

Given the large number of applications that rely on web APIs, it is essential to test these APIs thoroughly to ensure their quality. 
It is therefore not surprising that many automated techniques and tools for REST APIs have been proposed in recent years~\cite{arcuri2017restful,arcuri2019restful, atlidakis2019restler, godefroid2020intelligent, martin2020restest, karlsson2020quickrest, viglianisi2020resttestgen, corradiniautomated, atlidakis2020checking, atlidakis2020pythia, laranjeiro2021black, segura2017metamorphic, ed2018automatic, stallenberg2021improving}. 
These techniques take as input a description of the API, in the OpenAPI specification format~\cite{openapi} or API Blueprint~\cite{apiblueprint}, and employ various strategies to generate test cases for exercising API endpoints defined in the specification.

Although these tools have been evaluated, these evaluations have been performed (1) in different settings (in terms of API benchmarks considered, experiment setup, and metrics used), (2) using benchmarks that are in some cases limited in size or closed-source, and (3) mostly in isolation or involving only limited comparisons. 
It is thus difficult to understand how these tools compare to one another.
Recently, and concurrently to our effort, there has been some progress in this direction, via two empirical comparisons of black-box REST API testing tools~\cite{corradini2021empirical, Zac2021schemathesis} and an evaluation of white-box and black-box REST API test generation~\cite{martinlopez2021comparing}. 
These efforts are a step in the right direction but are still limited in scope.
Specifically, and to the best of our knowledge, ours is the first study that (1) systematically identifies both academic and practitioners' tools to be used in the comparison, (2) analyzes the code of the benchmarks to identify code characteristics that affect the performance of the tools and differentiate them, (3) conducts an in-depth analysis of the failures revealed by the tools, and (4) identifies concrete and specific future research directions.

As benchmark for our study, we used a set of 20 RESTful services selected among those used in related work and through a search on GitHub, focusing on Java/Kotlin open-source services that did not excessively rely on external resources. 
To select the tools for our evaluation, we performed a thorough literature search, which resulted in 8 academic and 11 practitioners' tools. 
Among those, we selected the tools that relied on commonly used REST API standards, such as OpenAPI, produced actual test cases, and were able to handle our benchmark of 20 services.
The resulting set consisted of 10 tools overall: one white-box tool, EvoMasterWB~\cite{arcuri2017restful, arcuri2019restful}, and nine black-box tools, APIFuzzer~\cite{apifuzzer}, bBOXRT~\cite{laranjeiro2021black}, Dredd~\cite{dredd}, EvoMasterBB~\cite{arcuri2020automated}, RESTest~\cite{martin2020restest}, RESTler~\cite{atlidakis2019restler}, RestTestGen~\cite{viglianisi2020resttestgen,corradiniautomated}, Schemathesis~\cite{Zac2021schemathesis}, and Tcases~\cite{tcases}.
In the paper, we provide a characterization of these tools along several dimensions, including their underlying input-generation approach, their support for stateful API testing, and the types of test oracles they use.

We applied these tools to our benchmark of 20 services and evaluated their performance in terms of code coverage achieved (lines, branches, and methods exercised) and different kinds of fault-detection ability (generic errors, unique failure points, and unique library failure points).
Through a thorough analysis of the results, we also investigated the strengths and weaknesses of the tools and of their underlying test-generation strategies.

Overall, all tools achieved relatively low line and branch coverage on many benchmarks, which indicates that there is considerable room for improvement.
Two common limitations of many tools, in particular, involve the inability of (1) generating input values that satisfy specific constraints (e.g., parameters that must have a given format), and (2) satisfying dependencies among requests  (e.g., this endpoint must be called before these other endpoints).
In general, we found that accounting for dependencies among endpoints is key to performing effective REST API testing, but existing techniques either do not consider these dependencies or use weak heuristics to infer them, which limits their overall effectiveness. 

The paper also discusses lessons learned and implications for future research based on our results.
For example, REST API testing techniques should leverage information embedded in the specification and the server logs to improve the quality of the test input parameters they generate.
For another example, dependencies among services could be detected through static analysis, if the source code is available, and through natural language processing techniques applied to the service specification.
Finally, in addition to discussing possible ways to improve REST testing approaches, we also present a proof-of-concept evaluation of these ideas that shows the feasibility of the suggested directions.

In summary, this work provides the following contributions:

\squishlist
  
\item A comprehensive empirical study of automated test-generation tools for REST APIs that involves 10 academic and practitioners' tools and assesses their performance on 20 benchmarks in terms of code coverage and different types of fault-detection ability.

\item An analysis of the strengths and weaknesses of the tools considered and their underlying techniques, with suggestions for improvement and a discussion of the implications for future research.

\item An artifact with the tools and benchmarks that can allow other researchers to replicate our work and build upon it~\cite{artifact}.

\squishend

\vspace{-8pt}
\section{Background}
\label{sec:background}


\subsection{REST APIs}

REST APIs are web APIs that follow the RESTful architectural style~\cite{fielding2000architectural}. Clients can communicate with web services through their REST APIs by sending HTTP requests to the services and receiving responses. 
Clients send requests to access and/or manipulate resources managed by the service, where a \textit{resource} represents data that a client may want to create, delete, or access. The request is sent to an API \textit{endpoint}, which is identified by a resource path, together with an \textit{HTTP method} that specifies the action to be performed on the resource.
The most commonly used methods are \textsc{post}, \textsc{get}, \textsc{put}, and \textsc{delete}, which are used to create, read, update, and delete a resource, respectively. 
The combination of an endpoint plus an HTTP method is called an \textit{operation}.
In addition to specifying an operation, a request can optionally also specify HTTP headers containing metadata (e.g., the data format of the resource targeted) and a body that contains the payload for the request (e.g., text in JSON format containing the input values). 

After receiving and processing a request, the web service returns a response that includes, in addition to headers and possibly a body, an HTTP status code---a three-digit number that shows the outcome of the request. Status codes are organized into five suitably numbered groups. 1xx codes are used for provisional purposes, indicating the ongoing processing of a request.
2xx codes indicate successful processing of a request. 
3xx codes indicate redirection and show, for instance, that the target resource has moved to a new URL (code 301). 
4xx codes indicate client errors. For example, a 404 code indicates that the client has requested a resource that does not exist. 
Finally, 5xx codes indicate server errors in processing the request. Among this group, in particular, the 500 code indicates an Internal Server Error and typically corresponds to cases in which the service contains a bug and failed to process a request correctly. For this reason, many empirical studies of REST API testing tools report the number of 500 status codes triggered as the bugs they found (e.g., ~\cite{arcuri2017restful, atlidakis2019restler, corradiniautomated,karlsson2020quickrest,martin2020restest,
  atlidakis2020pythia,ed2018automatic,Zac2021schemathesis}).

\subsection{OpenAPI Specifications}

The description of a service interface is typically provided by means of a specification that lists the available operations, as well as the input and output data types and possible return codes for each operation. A common, standard format for such specifications is the OpenAPI Specification~\cite{openapi} (previously known as Swagger).
Other examples of specification languages include RAML~\cite{raml} and API
Blueprint~\cite{apiblueprint}.

\begin{figure}[t]
\centering
\setlength\multicolsep{0pt}
\setlength\columnsep{3pt}
\begin{lstlisting}[multicols=2]
"/products/{productName}": {
 "get": {
  "operationId":
     "getProductByName",
  "produces":
     ["application/json"],
  "parameters": [{
    "name": "productName",
    "in": "path",
    "required": true,
    "type": "string"
   }],
   "responses": {
    "200": {
     "description":
       "successful operation",
     "schema": {
      "$ref":
        "#/definitions/Product"
    },
    "headers": {}
}}}}
\end{lstlisting}
\vspace*{-3pt}
\caption{An example OpenAPI specification.}
\vspace*{-15pt}
\label{openapi_example}
\end{figure}

\begin{table*}[t]
\centering \footnotesize
\caption{Overview of REST API testing techniques and tools.}
\vspace*{-8pt}
\label{tool_overview_table}
\begin{threeparttable}
\resizebox{\textwidth}{!}{
\begin{tabular}{lcclcllc}
\toprule
\textbf{Name} & \textbf{Website} & \textbf{Testing Approach} & \textbf{Test-Generation Technique}
& \textbf{Stateful} & \textbf{Oracle} & \textbf{Parameter Generation} & \textbf{Version Used}\\
\midrule
EvoMasterWB     &~\cite{evomastertool} & White-box & Evolutionary algorithm            & Yes
& Status code                                           & Random, Mutation-based, and Dynamic
& v1.3.0$^\S$ \\ 

EvoMasterBB     &~\cite{evomastertool} & Black-box & 
Random Testing                    & Yes
& Status code                                           & Random
& v1.3.0$^\S$ \\ 

RESTler       &~\cite{restlertool} & Black-box & Dependency-based algorithm        & Yes
& Status code and predefined checkers                  & Dictionary-based and Dynamic
& v8.3.0 \\ 

RestTestGen   &~\cite{corradiniautomated}$^\ddag$ & Black-box & Dependency-based algorithm        & Yes
& Status code and response validation                   & Mutation-based, Default, Example-based, Random, and Dynamic
& v2.0.0 \\ 

RESTest       &~\cite{restesttool} & Black-box & Model-based testing        & No
& Status code and response validation                   & Constraint-solving-based, Random, and Mutation-based
& Commit 625b80e \\ 

Schemathesis  &~\cite{schemathesis}  & Black-box & Property-based testing             & Yes
& Status code and response validation                   & Random and Example-based
& v3.12.3 \\ 

Dredd         &~\cite{dredd} & Black-box & Sample-value-based testing             & No
& Status code and response validation                    & Example-based, Enum-based, Default, and Dummy
& v14.1.0 \\ 

Tcases        &~\cite{tcases} & Black-box & Model-based testing                & No
& Status code                   & Random or Example-based
& v3.7.1 \\ 

bBOXRT        &~\cite{bboxrttool} & Black-box & Robustness testing             & No
& Status code and behavioral analysis     & Random and Mutation-based
& Commit 7c894247 \\ 

APIFuzzer     &~\cite{apifuzzer} & Black-box & Random-mutation-based testing       & No
& Status code     & Random and Mutation-based
& Commit e2b536f \\ 


\bottomrule
\end{tabular}}
\begin{tablenotes}
\item[$\ddag$] We obtained this tool directly from its authors when it was a private tool, so our version is not in the website.
\vspace*{-2pt}
\end{tablenotes}
\end{threeparttable}
\vspace*{-5pt}
\end{table*}

Figure~\ref{openapi_example} shows a fragment of the OpenAPI specification for
Features-Service, one of the benchmarks in our study. It specifies an API
endpoint \texttt{\small /products/\{productName\}}
(line~1), that supports HTTP method \textsc{get} (line~2). 
The id for this operation is \texttt{\small getProductByName}
(lines~3--4), and the format of the returned data is JSON (lines~5--6). 
To exercise this endpoint, the client \textit{must} provide a value for the required parameter
\texttt{\small productName} in the form of a path parameter (lines~7--12)
(e.g., a request \texttt{\small GET /products/p} where \texttt{\small p}
is the product name). 
If a product with that name exists and the service can
process the request correctly, the client would receive a response with
status code \texttt{\small 200} (line~14) and the requested
product data.
Lines~18--19 reference the definition of the product data type, which contains information about the product, such as its id, name, features, and constraints (we omit the definition for lack of space). 

Two operations have a \textit{producer-consumer} dependency relationship when one of the operations (producer) can return data needed as input by the other operation (consumer). For example, operation \texttt{\small GET /products/\{productName\}} has producer-consumer relationship with operation \texttt{\small DELETE /products/\{product\-Name\}/constraints/\{con\-straintId\}} (not shown in the figure). This is because a  request sent to the former can lead to a response containing a constraint id needed to make a request to the latter.

\section{Testing Tools Used In The Study}
\label{sec:tools}


\subsection{Tools Selection}

For selecting tools for the study, we searched for REST-API-related papers published since 2011 in top venues in the areas of software engineering, security, and web technologies (e.g., ASE, CCS, FSE, ICSE, ISSTA, NDSS, S\&P, TOSEM, TSE, USENIX, WWW, and WSDM).\footnote{The latest search was performed in December 2021.}
We identified relevant papers via keyword search, using the terms ``rest'', ``api'', and ``test'' and by transitively following citations in the identified papers.
Among the resulting set of papers, we selected those about techniques and tools that (1) rely on well-known REST API standards (i.e., OpenAPI, RAML, and API Blueprint), and (2) produce actual test cases.

This process resulted in 19 publicly-available tools: eight research tools---EvoMasterWB, Evo\-MasterBB, RESTler, RESTest, RestTestGen, bBOXRT, Schema\-thesis, and api-tester---and 11 practitioners' tools---fuzz-lightyear, fuzzy-swagger, swagger-fuzzer, APIFuzzer, TnT-Fuzzer, RESTApiTester, Tcases, gadolinium, restFuzzer, Dredd, and kotlin-test-client.\footnote{We define as research tools those whose technical details are presented in research papers.}
More recently, we found a technical report~\cite{Zac2021schemathesis} that describes two additional tools, cats~\cite{cats} and got-swag~\cite{gotswag}. Because Schemathesis, which is included in our study, outperforms them significantly~\cite{Zac2021schemathesis}, we did not include these tools in our comparison.
We then eliminated tools that either did not work at all or failed to run on our benchmark of 20 services. It is worth noting that the remaining 10 tools, which are listed in Table~\ref{tool_overview_table}, were also the most popular based on stars, watches, and forks in their GitHub repositories. 

\subsection{Tools Description} 

\sloppy
Table~\ref{tool_overview_table} presents an overview of the 10 tools we selected
along several dimensions: the URL where the tool is available (column~2); whether the tool is white-box or black-box (column~3);  test-generation strategy used by the tool (column~4); whether the tool produces test cases that exercise APIs in a stateful manner (column~5); the types of oracle used by the tool (column~6); the approach used by the tool to generate input parameter values (column~7); and the version of the tool (column~8).


\textbf{EvoMaster}~\cite{arcuri2018evomaster} can test a REST API in either white-box or black-box mode. In the study, we used the tool in both modes. We refer to the tool in the black-box mode as \textit{EvoMasterBB} and in the white-box mode as \textit{EvoMasterWB}. Given a REST API and the OpenAPI specification, both tools begin by parsing the specification to obtain the information needed for testing each operation. 
EvoMasterBB performs random input generation: for each operation, it produces requests to test the operation with random values assigned to its input parameters. EvoMasterWB requires access to the source code of the API. It leverages an evolutionary algorithm (the MIO algorithm~\cite{DBLP:journals/corr/abs-1901-01541} by default) to produce test cases with the goal of maximizing code coverage. Specifically, for each target (uncovered) branch, it evolves a population of tests by generating new ones while removing those that are the least promising (i.e., have the lowest fitness value) for exercising the branch in each iteration until the branch is exercised or a time limit is reached. 
EvoMasterWB generates new tests through sampling or mutation. The former produces a test from scratch by either
randomly choosing a number of operations and assigning random values to their input parameters (i.e., random sampling) or accounting for operation dependencies to produce stateful requests (i.e., smart sampling). The
approach based on mutation, conversely, produces a new test by changing either the structure
of an existing test or the request parameter values. EvoMasterBB and EvoMasterWB use an automated oracle that checks for service failures resulting in a 5xx status code.  Recent
extensions of the technique further improve the testing effectiveness by accounting for database states~\cite{DBLP:conf/gecco/ArcuriG19} and making better use of resources and dependencies~\cite{zhang2021resource}.

\textbf{RESTler}~\cite{atlidakis2019restler} is a black-box technique that
produces stateful test cases to exercise ``deep'' states of the target service.
To achieve this, RESTler first parses the input OpenAPI specification and infers
producer-consumer dependencies between operations. It then uses a search-based algorithm to produce sequences of requests that conform to the inferred dependencies. Each time a request is appended to a sequence, RESTler
executes the new sequence to check its validity. It leverages dynamic feedback
from such execution to prune the search space (e.g., to avoid regenerating
invalid sequences that were previously observed).  An early version of RESTler relies on a
predefined dictionary for input generation and targets finding 5xx failures.
Newer versions of the technique adopt more intelligent fuzzing
strategies for input generation~\cite{godefroid2020intelligent} and use
additional security-related checkers~\cite{atlidakis2020checking}.

\textbf{RestTestGen}~\cite{viglianisi2020resttestgen,corradiniautomated} is another black-box
technique that exploits the data dependencies between operations to produce test cases. 
First, to identify dependencies, RestTestGen matches names of input and output fields of different operations.
It then leverages the inferred dependency information, as well as predefined priorities of HTTP methods, to compute a testing order of operations and produce tests.  For each operation, RestTestGen produces two types of tests: nominal and error tests. 
RestTestGen produces nominal tests by assigning to the input parameters of an operation either (1) values dynamically obtained from previous responses (with high probability) or (2) values randomly generated, values provided as default, or example values (with low probability).
It produces error tests by mutating nominal tests to make them invalid (e.g., by removing a required
parameter). 
RestTestGen uses two types of oracles that check (1) whether the status code of a response is expected (e.g., 4xx for an error test) and (2) whether a response is syntactically compliant with the response schema defined in the API specification.

\textbf{RESTest}~\cite{martin2020restest} is a model-based black-box input generation technique that accounts for inter-parameter dependencies~\cite{martin2019catalogue}. 
An \textit{inter-parameter dependency} specifies a constraint among parameters in an operation that must be satisfied to produce a valid request (e.g., if parameter A is used, parameter B should also be used).  To produce test cases, RESTest takes as input
an OpenAPI specification with the inter-parameter dependencies specified in a
domain-specific language~\cite{martinlopez2021} and transforms such dependencies into constraints~\cite{idlreasoner}.
RESTest leverages constraint-solving and random input generation to produce nominal
tests that satisfy the specified dependencies and may lead to a successful (2xx)
response. It also produces faulty tests that either violate the dependencies or are not
syntactically valid (e.g., they are missing a required parameter). 
RESTest uses different types of oracles that check (1) whether the status code is different from 5xx, (2) whether a response is compliant with its defined schema, and (3) whether expected status codes are obtained for different
types of tests (nominal or faulty).

\textbf{Schemathesis}~\cite{Zac2021schemathesis} is a black-box tool that
performs property-based testing (using the Hypothesis library~\cite{hypothesis}).
It performs negative testing and defines five types of oracles to determine whether the
response is compliant with its defined schema based on status code, content
type, headers, and body payload. By default, Schemathesis produces non-stateful
requests with random values generated in various combinations and assigned to the input parameters. 
It can also leverage user-provided input parameter examples. 
If the API specification contains ``link'' values that specify operation dependencies, the tool can produce stateful tests as sequences of requests that follow these dependencies. 

\textbf{Dredd}~\cite{dredd} is another open-source black-box tool that validates
responses based on status codes, headers, and body payloads (using the Gavel
library~\cite{gavel}).  For input generation, Dredd uses sample values provided
in the specification (e.g., examples, default values, and enum values) and
dummy values.

\textbf{Tcases}~\cite{tcases} is a black-box tool that performs model-based testing.
First, it takes as input an OpenAPI specification and automatically constructs a model of the input space that 
contains key characteristics of the input parameters specified for each operation (e.g., a characteristic for an integer parameter may specify that its value is negative).
Next, Tcases performs \textit{each-choice} testing (i.e., 1-way combinatorial testing~\cite{kuhn2013introduction}) to produce test cases that ensure that a valid value is covered for each input characteristic.
Alternatively, Tcases can also construct an example-based model by
analyzing the samples provided in the specification and produce test cases using
the sample data only. For test validation, Tcases checks the status code based on the request validity (as determined by the model).

\textbf{bBOXRT}~\cite{laranjeiro2021black} is a black-box tool that aims to
detect robustness issues of REST APIs.  Given a REST API and its OpenAPI
specification, the tool produces two types of inputs, valid and invalid, for
robustness testing.  It first uses a random approach to find a set of valid
inputs whose execution can result in a successful (i.e., 2xx) status code. Next, it
produces invalid inputs by mutating the valid inputs based on a predefined set
of mutation rules and exercises the REST API with those inputs. 
The approach involves manual effort to analyze the service behavior and (1) classify it based on a failure mode scale (the CRASH scale~\cite{koopman1997comparing}) and (2) assign a set of behavior tags that provide diagnostic information.

\textbf{APIFuzzer}~\cite{apifuzzer} is a black-box tool that performs fuzzing of REST APIs. Given a REST API and its specification, APIFuzzer first parses the specification to identify each operation and its properties. Then, it generates random requests conforming to the specification to test each operation and log its status code.
Its input-generation process targets the body, query string, path parameter, and headers of requests and applies random generation and mutation to these values to obtain inputs.
APIFuzzer uses the generated inputs to submit requests to the target API and produces test reports in the JUnit XML format.

\section{Empirical Study}
\label{sec:empirical_study}

In our study, we investigated three research questions for the set of tools we considered and described in the previous section:

\squishstart

\item \textbf{RQ1}: How much code coverage can the tools achieve?

\item \textbf{RQ2}: How many error responses can the tools trigger?

\item \textbf{RQ3}: What are the implications of our findings?

\squishend

To answer RQ1, we evaluated the tools in terms of the line, branch, and method coverage they achieve to investigate their abilities in exercising various parts of the service code.

For RQ2, we compared the number of 500 errors triggered by the tools to investigate their fault-finding ability, as is commonly done in empirical evaluations of REST API testing techniques (e.g.,~\cite{arcuri2017restful, atlidakis2019restler, viglianisi2020resttestgen, karlsson2020quickrest, martin2020restest, atlidakis2020pythia, ed2018automatic, Zac2021schemathesis, corradiniautomated, wu2022combinatorial, liu2022morest}). 
We measured 500 errors in three ways. 
First, to avoid counting the same error twice, we grouped errors by their stack traces, and reported \textit{unique 500 errors}. Therefore, unless otherwise noted, 500 errors will be used to denote unique 500 errors in the rest of the paper. 
Second, different unique 500 errors can have the same failure point (i.e., the method-line pair at the top of the stack trace). Therefore, to gain insight into occurrences of such failure sources, we also measured \textit{unique failure points}, which group stack traces by their top-most entries. 
Finally, we differentiated cases in which the unique failure point was a method from the web service itself from cases in which it was a third-party library used by the service; we refer to the latter as \textit{unique library failure points}.

To answer RQ3, we (1) provide an analytical assessment of the studied tools in terms of their strengths and
weaknesses and (2) discuss the implications of our study for the development of new techniques and tools in this space.


Next, we describe the benchmark of web services we considered (\S\ref{sec:benchmark}) and our experiment setup
(\S\ref{sec:experiment-setup}). We then present our results for the three research questions (\S\ref{sec:coverage-result}--\ref{sec:implication}). We conclude this section with a discussion of potential threats to validity of our results (\S\ref{sec:threats}).

\subsection{Web Services Benchmark}
\label{sec:benchmark}

\begin{table}[t]
\centering \footnotesize
\tabcolsep=3pt
\caption{RESTful web services used in the empirical study.}
\vspace*{-7pt}
\label{overview-benchmarks}
\begin{tabular}{lccc} 
\toprule
\textbf{Name} & \textbf{Total LOC} & \textbf{Java/Kotlin LOC} & \textbf{\# Operations}\\ 
\midrule
CWA-verification & 6,625 & 3,911 & 5\\ 
ERC-20 RESTful service & 1,683 & 1,211 & 13\\ 
Features-Service & 1,646 & 1,492 & 18 \\ 
Genome Nexus & 37,276 & 22,143 & 23\\ 
Languagetool & 677,521 & 113,277 & 2\\ 
Market & 14,274 & 7,945 & 13\\ 
NCS & 569 & 500 & 6\\ 
News & 590 & 510 & 7\\ 
OCVN & 59,577 & 28,093 & 192\\ 
Person Controller & 1,386 & 601 & 12\\ 
Problem Controller & 1,928 & 983 & 8\\ 
Project tracking system & 88,634 & 3,769 & 67\\ 
ProxyPrint & 6,263 & 6,037 & 117\\ 
RESTful web service study & 3,544 & 715 & 68\\ 
Restcountries & 32,494 & 1,619 & 22\\ 
SCS & 634 & 586 & 11\\ 
Scout-API & 31,523 & 7,669 & 49\\ 
Spring boot sample app & 1,535 & 606 & 14\\ 
Spring-batch-rest & 4,486 & 3,064 & 5\\ 
User management & 5,108 & 2,878 & 23\\ 
\midrule
Total & 977,296 & 207,609 & 675\\ 
\bottomrule
\end{tabular}
\vspace*{-8pt}
\end{table}

To create our evaluation benchmark, we first selected RESTful web services from existing benchmarks used in the evaluations of bBOXRT, EvoMaster, and RESTest. 
Among those, we focused on web services implemented in Java/Kotlin and available as open-source projects, so that we could use EvoMasterWB. This resulted in an initial set of 30 web services. 
We then searched for Java/Kotlin repositories on GitHub using tags ``rest-api'' and ``restful-api'' and keywords ``REST service'' and ``RESTful service'', ranked them by number of stars received, selected the 120 top entries, and eliminated those that did not contain a RESTful specification. 
This additional search resulted in 49 additional web services. 
We tried installing and testing this total set of 79 web services locally and eliminated those that (1) did not have OpenAPI Specifications, (2) did not compile or crashed consistently, or (3) had a majority of operations that relied on external services with request limits (which would have unnaturally limited the performance of the tools).
For some services, the execution of the service through the EvoMaster driver class that we created crashed without a detailed error message.
Because this issue was not affecting a large number of services, we decided to simply exclude the services affected.

In the end, this process yielded 20 web services, which are listed in Table~\ref{overview-benchmarks}.
For each web service considered, the table lists its name, total size, size of the Java/Kotlin code, and number of operations.
We obtained the API specifications for these services directly from their corresponding  GitHub repositories. 

\subsection{Experiment Setup}
\label{sec:experiment-setup}

We ran our experiments on Google Cloud e2-standard-4 machines running Ubuntu 20.04. Each machine had 4 2.2GHz Intel-Xeon processors and 16GB RAM. We
installed on each machine the 10 testing tools from Table~\ref{tool_overview_table}, the 20 web services listed in Table~\ref{overview-benchmarks}, and other software used by the tools and services, including OpenJDK~8, OpenJDK~11, Python~3.8, Node.js~10.19, and .NET framework~5.0. We also set up additional services used by the benchmarks, such as MySQL, MongoDB, and a private Ethereum network. This process was performed by creating an image with all the tools, services, and software installed and then instantiating the image on the cloud machines.

We ran each tool with the recommended configuration settings as described in the respective paper and/or user manual. Some of the tools required additional artifacts to be created (see also Section~\ref{sec:threats}). For RESTest, we created inter-parameter dependency specifications for the benchmark applications, and for EvoMasterWB, we implemented driver programs for instrumenting Java/Kotlin code and database injection. 
For lack of space, we omit here details of tool set up and configuration. A comprehensive description that enables our experiments and results to be replicated is available at our companion website~\cite{artifact}.

We ran each tool for one hour, with 10 trials to account for randomness.
Additionally, we ran each tool once for 24 hours, using a fresh machine for each run to avoid any interference between runs.
For the one-hour runs, we collected coverage and error-detection data in 10-minute increments to investigate how these metrics increased over time. 
For the 24-hour run, we collected the data once at the end. We collected coverage information using the JaCoCo code-coverage library~\cite{jacoco}.

\subsection{RQ1: Coverage Achieved}
\label{sec:coverage-result}

\begin{table}[t]
\centering \footnotesize
\caption{Average line, branch, and method coverage achieved and 500 errors found in one hour (E: unique error, UFP: unique failure point, ULFP: unique library failure point).}
\vspace*{-7pt}
\label{test_result_table}
\begin{tabular}{lcccc} 
\toprule
\textbf{Tool} & \textbf{Line} & \textbf{Branch} & \textbf{Method} & \textbf{\#500 (E/UFP/ULFP)} \\ 
\midrule
EvoMasterWB & \textbf{52.76\%}   & \textbf{36.08\%}   & \textbf{52.86\%} & \textbf{33.3} / \textbf{6.4} / \textbf{3.2} \\ 
RESTler    & 35.44\%   & 12.52\%   & 40.03\% & 15.1 / 2.1 / 1.3  \\ 
RestTestGen & 40.86\%  & 21.15\%   & 42.31\% & 7.7 / 2 / 1  \\ 
RESTest   & 33.86\%   & 18.26\%  & 33.99\% & 7.2 / 1.9 / 1.1 \\ 
bBOXRT & 40.23\% & 22.20\% & 42.48\% & 9.5 / 2.1 / 1.3\\ 
Schemathesis & 43.01\%   & 25.29\%   & 43.65\% & 14.2 / 2.8 / 2 \\ 
Tcases & 37.16\% & 16.29\%   & 41.47\% & 18.5 / 3.5 / 2.1 \\ 
Dredd & 36.04\% & 13.80\% & 40.59\% & 6.9 / 1.5 / 0.9 \\ 
EvoMasterBB & 45.41\%   & 28.21\%   & 47.17\% & 16.4 / 3.3 / 1.8  \\ 
APIFuzzer & 32.19\% & 18.63\% & 33.77\% & 6.9 / 2.2 / 1.3\\ 
\bottomrule
\end{tabular}
\vspace*{-10pt}
\end{table}
\begin{figure*}[!t]
    \includegraphics[width=\linewidth]{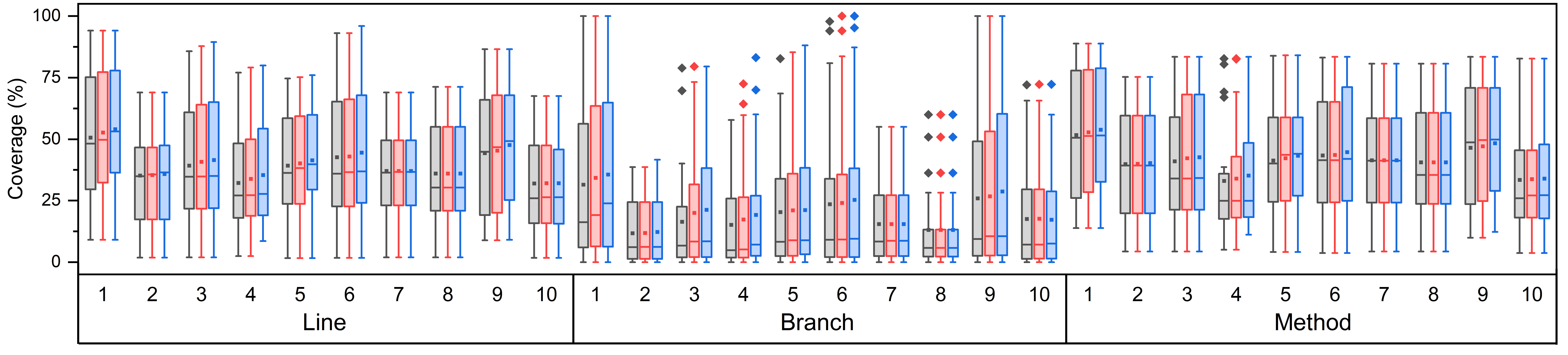}
    \vspace*{8pt}
    \includegraphics[width=\linewidth]{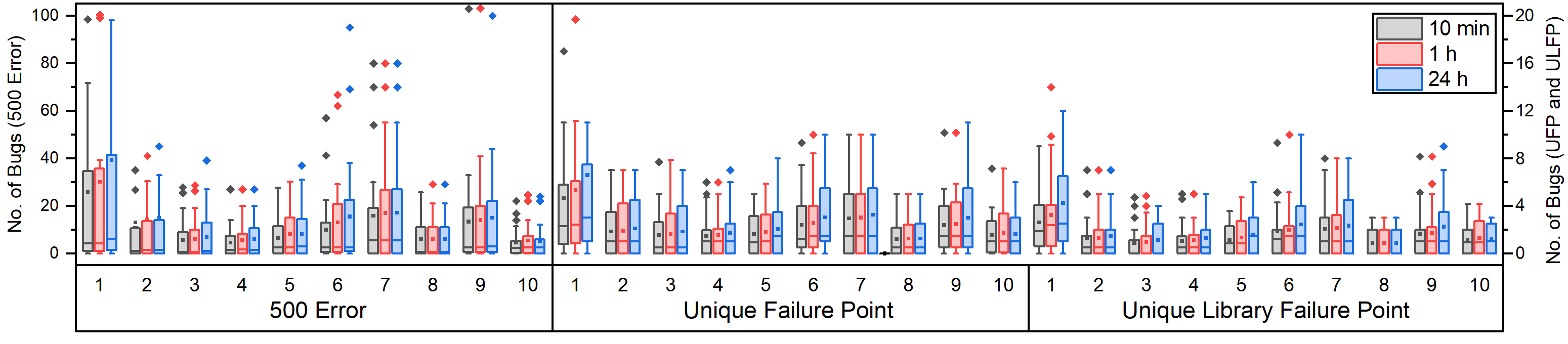}
    \vspace*{-18pt}
    \caption{Code coverage achieved and number of unique 500 errors, unique failure points, and unique library failure points detected over all services by the ten tools in 10 minutes, 1 hour, and 24 hours (1: EvoMasterWB, 2: RESTler, 3: RestTestGen, 4: RESTest, 5: bBOXRT , 6: Schemathesis, 7: Tcases, 8: Dredd, 9: EvoMasterBB, 10: APIFuzzer).}
    \label{fig:benchmark}
\end{figure*}

Table~\ref{test_result_table} presents the results for line, branch, and method coverage achieved, and 500 errors, unique failure points, and unique library failure points detected by the 10 tools in one hour. 
The reported results are the average of the coverage achieved and errors found over all services and all ten trials. 
Unlike Table~\ref{test_result_table}, which shows only average results, Figure~\ref{fig:benchmark} presents coverage data (top) and error-detection results (bottom) as box plots.
In the figure, the horizontal axis represents the 10 tools, and the vertical axis represents, for each tool, the results achieved by that tool after 10 minutes, 1 hour, and 24 hours. 
Each box represents average, median, standard deviation, min, and max for 20 data points, one per service considered. For the 10-minute and 1-hour runs, each point represents the coverage achieved (or number of errors revealed) averaged over 10 trials. For the 24-hour runs, the values reported are from single trials.

\paragraph{Overall coverage achieved}

Table~\ref{test_result_table} shows that the tools did not achieve a high coverage in general. In one hour, the best-performing tool (EvoMasterWB) achieved less than 53\% line coverage, less than 37\% branch coverage, and less
than 53\% method coverage. 
The other black-box tools achieved lower coverage; the best-performing tool among them is EvoMasterBB, with 45.41\% line coverage, 28.21\% branch coverage, and 47.17\% method coverage.  We identified three key factors responsible for these low coverage rates.

\squishstart

\item \textbf{Parameters value generation}. The tools generate many
  invalid requests that are rejected by the services and fail to exercise service
  functionality in depth. In most cases, this occurs when the parameters take domain-specific values or have data-format restrictions. As an example, for a
  parameter that takes email IDs as values, one of the services enforced the
   condition that the email should contain a `@' character. 
   Tools wasted a significant amount of time attempting to pass that check, but they all failed in the one-hour and 24-hour runs in our experiment.

\item \textbf{Operations dependency detection}. Identifying producer-consumer
  dependencies between operations is key to generating stateful tests. Existing
  tools either do not account for such dependencies or use a simple heuristic
  approach, which can lead to many false positives and false negatives in the
  dependencies computed. Failing to identify a dependency between producer A
  and consumer B can result in inadequate testing of B due to not testing A
  first. This is because B may require the execution of A to set the service
  state needed for its testing (e.g., by adding valid data to the database) or
  obtain valid parameter values. As an example, for Languagetool, most of the tools
  fail to identify a dependency between the two operations \texttt{\small
    POST /check} (consumer) and \texttt{\small GET /languages} (producer) in 24 hours. As a result, they did not leverage the output of the producer, which contained
  language information, and were unable to produce valid requests for the consumer.


\item \textbf{Mismatch between APIs and their specifications}. The tools produce test
cases by leveraging the API specifications, which are expected to faithfully reflect
the API implementations and define, for each endpoint, supported HTTP methods, 
required input parameters, responses, and so on. Unfortunately, this is not always the case and, if the specification does not match the API implementation, the tools can produce
incomplete, invalid, or even no requests at all for exercising the affected operations.

\commentt{
\saurabh{Keep this?}\myeongsoo{I think we can remove it. I just wanted to avoid the reviewers could complain that the incompletion is our benchmark dataset}
To investigate the prevalence of this issue, we sampled 1345 OpenAPI specifications from APIs guru~\cite{apisgurugithub} and ran Schemathesis on them to check whether actual service responses conform to the response schema defined in the specifications. We found that 98.77\% of the specifications has at least one response mismatch.
}

\squishend

\begin{tcolorbox}

Existing tools fail to achieve high code coverage due to limitations of the
approaches they use for generating parameter values and detecting operation
dependencies. 
The effectiveness of the tools is also hindered by mismatches between API implementations and specifications.

\end{tcolorbox}

\paragraph{Coverage increase over time}

As we mentioned above, we collected coverage data in 10-minute increments, as well as for 24 hours.
Figure~\ref{fig:benchmark} illustrates how the coverage rate increased from 10 minutes to 1~hour, and to 24 hours
(more detailed results are available in our artifact~\cite{artifact}).
As the figure shows, in many cases, the tools already achieved their highest level of code coverage in 10 minutes.
However, there were several cases in which code coverage kept increasing over time. 
We investigated the 6 services for which the tools manifested this behavior---Features-Service, NCS, Restcountries, SCS, RESTful Web Service Study, and User management---and found that they all have some  distinct characteristics (compared to the other 14 services).
Specifically these services generally have simpler input parameters that do not take domain-specific or otherwise constrained values, so additional input generation tends to cover additional code.
Conversely, for the services that have domain-specific or constrained parameter values (e.g., ``email should contain @'' or ``year should be between 1900 and 2200''), the tools tend to hit a coverage wall relatively soon because they are inherently unable to generate inputs that satisfy these requirements.

\begin{tcolorbox}
The coverage achieved by testing tools grows over time on services that have simpler input parameters with no domain-specific or constrained values.
\end{tcolorbox}

\subsection{RQ2: Error Responses Triggered}
\label{sec:finding-faults-result}

The ultimate goal of testing is to find bugs. In this section, we focus on comparing the testing tools in terms of their fault-finding ability, which we measured in terms of the numbers of unique 500 errors, failure points, and library failure points detected.
Column~5 in the previously presented Table~\ref{test_result_table} provides this information averaged
over the services considered and the trials performed. 
As the table shows, EvoMasterWB is the best performer by a wide margin, followed by Tcases, EvoMasterBB, RESTler, and Schemathesis. The box plot at the bottom of Figure~\ref{fig:benchmark} presents a more detailed view of these results by
illustrating the distribution of errors detected across services and
the increase in errors detected over time.
The ``500 Error'' segment of the plot shows that EvoMasterWB outperforms
all the other tools in terms of the median values as well, although with a larger
spread.

\begin{tcolorbox}
EvoMasterWB, by having access to source code and performing coverage-driven testing, achieves significantly higher coverage than black-box tools. However, as we will show in Section~\ref{sec:analytical-comparison}, there are cases in which EvoMasterWB cannot produce requests covering some service functionality, whereas black-box tools can.
\end{tcolorbox}

The figure also shows that the number of unique failure points is considerably smaller than the number of unique errors---on average, there are 3 to 7 times fewer failure points than errors, indicating that several 500 errors occur at the same program points but along different call chains to those points. 
Another observation is that approximately half of the unique failure points occur in library methods. A more detailed analysis of these cases revealed that failure points in library methods could
have more serious consequences on the functionality and robustness of a service, in some cases also leaving it vulnerable to security attacks.
The reason is that failures in service code mostly originate at statements that throw exceptions after performing some checks on input values; the
following fragment is an illustrative example, in which the thrown exception is automatically mapped to a 500 response code by the REST API framework being used:

\begin{lstlisting}
if (product == null) {
    throw new ObjectNotFoundException(name);
}
\end{lstlisting}

Conversely, in the case of library failure points, the root cause of the failures was often an unchecked parameter value, possibly erroneous, being passed to a library method, which could cause severe issues. 
We found a particularly relevant example in the ERC-20 RESTful service, which uses an
Ethereum~\cite{wood2014ethereum} network. An Ethereum transaction requires a fee, which is referred to as gas. 
If an invalid request, or a request with insufficient gas, is sent to Ethereum by the service, the transaction is canceled, but the gas fee is not returned.
This is apparently a well-known attack scenario in Blockchain~\cite{saad2019exploring}.
In this case, the lack of suitable checks in the ERC-20 service for requests sent
to Ethereum could have costly repercussions for the user. We found other examples
of unchecked values passed from the service under test to libraries, resulting
in database connection failures and parsing errors.

\begin{tcolorbox}
The severity of different 500 errors can vary considerably. Failure points in the service code usually occur at throw statements following checks on parameter values. Failure points outside the service, however, often involve erroneous requests that have been accepted and processed, which can lead to severe failures.
\end{tcolorbox}


\sloppy
We further investigated which factors influence the error-detection ability of the tools. 
First, we studied how the three types of coverage reported in Table~\ref{test_result_table} correlate with the numbers of the 500 errors found using the Pearson's Correlation Coefficient~\cite{freedman2007statistics}. 
The results showed that there is a strong positive correlation between the coverage and number of 500 errors
(coefficient score is $\bettersim 0.7881$ in all cases). Although expected, this result confirms that it makes sense for tools to use coverage as a goal for input generation: if a tool achieves higher coverage, it will likely trigger more failures. 

\begin{tcolorbox}

There is a strong positive correlation between code coverage and number of faults exposed. 
Tools that achieve higher coverage usually also trigger more failures.
\end{tcolorbox}

Second, we looked for patterns, not related to code coverage, that can increase the likelihood of failures being triggered. This investigation revealed that exercising operations with different combinations of parameters can be particularly effective in triggering service failures. 
The failures in such cases occur because the service lacks suitable responses for some parameter combinations. In fact, Tcases and Schemathesis apply this strategy to their advantage and outperform the other tools. Generating requests with different input types also seems to help in revealing more faults.

\begin{tcolorbox}
Exercising operations with various parameter combinations and various input types helps revealing more faults in the services under test.
\end{tcolorbox}

\subsection{RQ3: Implications of Our Results}
\label{sec:implication}

We next discuss the implications of our results for future research
on REST API testing and provide an analytical assessment of testing
strategies employed by the white-box and black-box testing tools we considered.

\begin{figure}
\centering
\begin{lstlisting}
public static String subject(String directory, String file) {
  int result = 0;
  String[] fileparts = null;
  int lastpart  = 0;
  String suffix = null;  
  fileparts = file.split(".");
  lastpart = fileparts.length - 1;
  if (lastpart > 0) { ... } //target branch
  return "" + result;
}
\end{lstlisting}
\vspace*{-8pt}
\caption{A method used for parsing the file suffix in the SCS web service.}
\vspace*{-12pt}
\label{file_suffix_code}
\end{figure}

\subsubsection{Implications for Techniques and Tools Development}

\paragraph{Better input parameter generation} Our results show that the
tools we considered failed to achieve high code coverage and could be considerably improved. 
Based on our findings, one promising direction in this context is better input parameter generation.

In particular, for white-box testing, analysis of source code can provide critical guidance on
input parameter generation. In fact, EvoMasterWB, by
performing its coverage-driven evolutionary approach, achieves higher coverage
and finds more 500 errors than any of the black-box tools. 
There are, however, situations
in which EvoMasterWB's approach cannot direct its input search toward better
coverage. Specifically, this happens when the fitness function used is ineffective
in computing a good search gradient. As an example, for the code shown in
Figure~\ref{file_suffix_code}, which parses the suffix of a file, EvoMasterWB
always provides a string input \texttt{\small file} that leads to \texttt{\small
  lastpart} in line~8 evaluating to~0. The problem is that EvoMasterWB cannot derive a gradient
from the condition \texttt{\small lastpart > 0} to guide the generation of
inputs that exercise the branch in line~8. In this case, symbolic
execution~\cite{king1976symbolic, baldoni2018survey} could help find a good
value of \texttt{\small file} by symbolically interpreting the method and
solving the constraint derived at line~8.

For black-box testing approaches, which cannot leverage information in the source code, using more sophisticated testing techniques (e.g., combinatorial testing at 2-way or higher interaction levels) could be a promising direction~\cite{wu2022combinatorial,karlsson2020automatic,vosta2020evaluation}. Also, black-box testing tools could try to leverage useful information embedded in the specification and other resources to guide input parameter generation.

Our analysis in Section~\ref{sec:analytical-comparison} below shows that using sample values for input
parameter generation can indeed lead to better tests. 
Therefore, another possible way to improve test generation would be to automatically extract sample values from parameter descriptions in the specification. 
For example, the description of the input parameter \texttt{\small language}, shown
in the following OpenAPI fragment for Languagetool (lines~4--9), suggests useful input
values, such as ``en-US'' and ``en-GB'':

\begin{lstlisting}
"name": "language",
"in": "formData",
"type": "string",
"description": "A language code like `en-US`, `de-DE`, `fr`, 
or `auto` to guess the language automatically 
(see `preferredVariants` below). For languages with variants 
(English, German, Portuguese) spell checking will
only be activated when you specify the variant, 
e.g. `en-GB` instead of just `en`.", 
"required": true
\end{lstlisting}

Another source of useful hints for input generation can be response messages from the server, such as the following one:

\begin{lstlisting}
Received: "HTTP/1.1 400 Bad Request\r\nDate: Wed, 28 Apr 2021 
00:38:32 GMT\r\nContent-length: 41\r\n\r\nError: Missing 'text' 
or 'data' parameter"
\end{lstlisting}

To investigate the feasibility of leveraging input parameter descriptions provided in the REST API specifications to obtain valid input values, we implemented a proof-of-concept text-processing tool that checks whether there are suggested values supplied in the parameter description for a given input parameter. We applied the tool to two of the services, OCVN and Languagetool, for which the existing testing tools failed to obtain high coverage (less than 10\% line code coverage for most cases). With the help of the tool, we identified developer-suggested values for 934 (32\%) of the 2,923 input parameters. This preliminary result suggests that leveraging parameter descriptions for input generation may be a feasible and promising direction.


Along similar lines, we believe that natural-language processing (NLP)~\cite{manning1999foundations} could be leveraged to analyze the parameter description and extract useful information. For example, a technique may first perform token matching~\cite{webster1992tokenization} to identify what parameters from the specification are mentioned in server messages and then use parts-of-speech tagging~\cite{voutilainen2003part} and dependency parsing~\cite{kubler2009dependency} to infer parameter values.

\begin{figure*}[t]
    \centering
    \includegraphics[width=.4\linewidth]{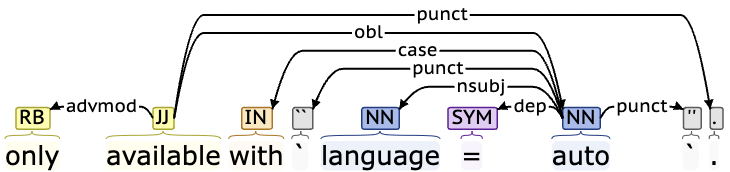}
    \vskip 6pt
    \includegraphics[width=\linewidth]{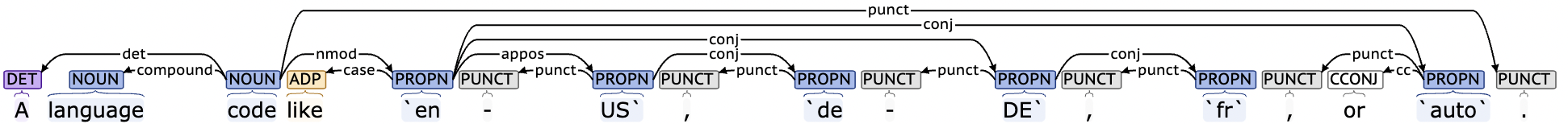}
    \vspace*{-16pt}
    \caption{Parsed dependency graphs of \texttt{\small preferredVariants} parameter description (top) and \texttt{\small language} parameter description
    (bottom) from Languagetool's OpenAPI specification.}
    \vspace*{-10pt}
    \label{fig:preferredVariants}
\end{figure*}

Also in this case, to investigate the feasibility of this approach, we implemented a
proof-of-concept prototype that parses dependency information from
natural-language descriptions in OpenAPI specifications and server messages and
collects nouns, pronouns, conjunctions, and their dependencies. 
Our prototype detects parameter names with simple string matching on nouns and pronouns, and
relationships between parameters via conjunctions and dependencies.
As an example, the top part of Figure~\ref{fig:preferredVariants} shows the parsed 
dependencies for parameter \texttt{\small preferredVariants} of endpoint
\texttt{\small /check}. By analyzing the tokens \texttt{\small with},
\texttt{\small language}, and \texttt{\small auto} with the connected
dependencies (case, punct, nsubj, and dep), our prototype can determine that
parameter \texttt{\small language}  must be set to ``auto''.
This simple approach was able to automatically detect 
8 of the 12 unique inter-parameter dependencies that we manually found
in the benchmark APIs. It was also able to detect useful parameter values.
For example, none of the black-box tools studied could generate
values for parameter \texttt{\small language}; yet, our prototype detected
useful values such as ``en-US'', ``de-DE'', and ``fr''
(the parsed dependencies are shown in the bottom part of Figure~\ref{fig:preferredVariants}).
As before, these preliminary results show the feasibility of applying NLP techniques
to descriptions in OpenAPI specifications and server messages for improving REST API testing.

\paragraph{Better support for stateful testing}

As discussed in Section~\ref{sec:coverage-result} and also stressed in related work~\cite{arcuri2018evomaster,atlidakis2019restler}, producing stateful tests by inferring producer-consumer relationships between operations is key to effective REST API testing.  
Current tools, such as RestTestGen and RESTler, rely on
simple heuristics to infer the producer-consumer dependencies and are inaccurate in identifying such dependencies, which can easily lead to false positives and negatives.  There is a need for more sophisticated approaches that address the flexibility issues that we discuss later in Section~\ref{sec:analytical-comparison}.  These approaches should be able to account for responses dynamically generated and to match fields of different types while still being precise (e.g., by not identifying a dependency between two consumers that rely on the same parameters to execute).

To determine whether two operations have a dependency relationship, one possible approach would be to check whether they have ``related'' properties, such as input parameters and response fields, that have the same or similar names. To explore the feasibility of this direction, we performed a quick case study in which we manually identified each operation that uses path parameters and has a dependency relationship with at least another operation. For each of them, we then identified its path parameter(s) and compared its textual similarity with other operations' response fields using the NLTK's scoring library~\cite{nltk}. In this way, by considering the top-3 matches for each parameter, we were able to correctly identify almost 80\% of the operations involved in a dependency relationship. Based on our findings in RQ1, we expect that a similar approach could considerably help tools to achieve higher code coverage.

We also believe that machine learning~\cite{jordan2015machine} could help identify producer-consumer relationships and related dependencies. For example, one could train, via supervised learning~\cite{zhu2009introduction}, a
classifier that accounts for a variety of features related to operation
dependencies (e.g., HTTP methods, shared input parameters, field matching on dynamic object, field types) and then use this classifier to make predictions on potential dependencies.

\subsubsection{Analytical Comparison of White-Box and Black-Box Tools}
\label{sec:analytical-comparison}
\hfill\\

Next, we present an analytical assessment of the tools with respect to
strengths and weaknesses of their approaches for generating input parameter
values and sequences of API requests. We also provide illustrative examples
taken from the benchmark services.

\begin{figure}[t]
\centering
\begin{lstlisting}
if ("male".equals(sex)) {
    if ("mr".equals(title) || "dr".equals(title) || 
        "sir".equals(title) || "rev".equals(title) ||
        "rthon".equals(title) || "prof".equals(title)) {
            result = 1;
    }
} else if ("female".equals(sex)) {
    if ("mrs".equals(title) || "miss".equals(title) ||
        "ms".equals(title) || "dr".equals(title) ||
        "lady".equals(title) || "rev".equals(title) ||
        "rthon".equals(title) || "prof".equals(title)) {
            result = 0;
}}
\end{lstlisting}
\vspace*{-12pt}
\caption{Sample code from the SCS web service.}
\vspace*{-16pt}
\label{sex_example}
\end{figure}

\paragraph{White-box vs. black-box tools}
Among the tools we considered, EvoMasterWB is the only one that performs white-box testing. By
having access to the API source code and performing coverage-driven testing,
EvoMasterWB achieves higher coverage than the other tools---according to
Table~\ref{test_result_table}, it achieves $\bettersim$53\%
line coverage, $\bettersim$36\% branch coverage, and $\bettersim$53\% method coverage.

To illustrate, Figure~\ref{sex_example} shows an if-statement in which the two branches
(lines~5 and 12) are exercised only by test cases produced by EvoMasterWB.  The
if-statement is responsible for handling requests for operation
\texttt{\small GET /api/title/\{sex\}/\{title\}} of the SCS web service.  To
cover these branches, a tool must produce valid requests with relevant string
values for parameters \texttt{\small sex} and \texttt{\small title}.  Using an
evolutionary algorithm with a fitness function that measures branch distance,
EvoMasterWB successfully generates values ``male'' for \texttt{\small sex} and
``dr'' for \texttt{\small title} to exercise line~5, and values ``female'' and
``prof'' for those parameters to cover line~12. The black-box tools are unable
to do this by using random and/or sample values.

Another benefit of EvoMasterWB's testing strategy is that once it produces a test
case exercising a branch A, it will not generate similar test cases to exercise
A again. This is due to its MIO algorithm~\cite{DBLP:journals/corr/abs-1901-01541}, which handles
test case generation for the target branches separately and focuses on uncovered branches.

\sloppy
However, there are also cases in which EvoMasterWB fails to create a sequence of 
requests for covering some API functionality, whereas a black-box tool is
able to. Consider the code fragment shown in
Figure~\ref{features-service_config_example}, which handles requests for
operation $O_2$: \texttt{\small GET /products/\{productName\}/configurations} of
Features-Service. To cover line~6, a request must specify a product that exists
and has configurations associated with it. None of the requests created by
EvoMasterWB satisfy both conditions: operation $O_1$: \texttt{\small POST
  /products/\{productName\}/configurations/\{configurationName\}} must be called
before $O_2$ to associate configurations with a product, and EvoMasterWB fails to
recognize this producer-consumer relation.  In contrast, Schemathesis uses a
testing strategy that orders $O_1$ before $O_2$ and leverages sample values from
the API specification to link two operations with the same input
values. It can therefore generate a sequence that creates a product, adds configurations to
it, and retrieves the configurations---a sequence that covers line~6.

\commentt{
\begin{tcolorbox}
EvoMasterWB, by having access to source code and performing coverage-driven
testing, achieves significantly higher coverage than black-box tools. However,
there are cases in which EvoMasterWB cannot produce requests covering some service
functionality, whereas a black-box tool can.
\end{tcolorbox}
}

\begin{figure}
\centering
\begin{lstlisting}
public List<String> 
  getConfigurationsNamesForProduct(String productName) {
  List<String> configurationsForProduct=new ArrayList<String>();
  for (ProductConfiguration productConfiguration : 
    productsConfigurationsDAO.findByProductName(productName)) {
    configurationsForProduct.add(productConfiguration.getName());
  }
  return configurationsForProduct;
}
\end{lstlisting}
\vspace*{-10pt}
\caption{A method of Features-Service used to find configuration names associated with a product.}
\vspace*{-12pt}
\label{features-service_config_example}
\end{figure}

\paragraph{Assessment of black-box tools}

Although black-box tools seems to be less effective than EvoMasterWB in terms of coverage achieved, which can also often result in fewer faults triggered, they have wider applicability by virtue of being agnostic to the language used for the service implementation. Among the black-box tools, EvoMasterBB and Schemathesis achieve better coverage than the other tools in terms of all the three metrics considered (unique 500 errors, failure points, and library failure points).

EvoMasterBB uses an evolutionary algorithm to generate successful requests, create sequences from them, and discover in this way operation dependencies. This way of operating allows it to find more valid request sequences than other tools that just use a randomized approach.

Schemathesis is the next best black-box tools. One characteristic of this tool is that it considers example values provided in the API specification, which lets it leverage the domain knowledge encoded in such examples. Moreover, it reuses values in creating request sequences, which enables it to successfully create covering sequences such as that for the loop body in Figure~~\ref{features-service_config_example}.

RestTestGen also has similar features, but we empirically found that its heuristic algorithm, which relies upon the matching of response fields and parameter names to infer producer-consumer dependencies between operations, yields many false positives and false negatives. This issue weakens the effectiveness of its stateful testing and leads to slightly lower code coverage than EvoMasterBB.

RESTler tries to infer producer-consumer relationships by leveraging feedback from processed requests. An important limiting factor for RESTler, however, is that it relies on a small set of dictionary values for input generation, which hinders its ability to exercise a variety of service behaviors.

Dredd does not perform random input generation but uses dummy values and input examples (provided in the specification) to produce requests.  This prevents Dredd from generating invalid requests exercising uncommon behaviors in a service. 

Finally, the other tools' random-based approaches are unlikely to produce valid test cases needed for input mutation, which limits their overall effectiveness.

Overall, black-box tools fail to achieve high coverage because they largely rely on random
testing and/or leverage a limited set of sample data for input generation.
Among these tools, Schemathesis and EvoMasterBB performed better than the others in our experiments due to some specific characteristics of the input generation approaches they use.

\subsection{Threats to Validity}
\label{sec:threats}


Like any empirical evaluation, our study could suffer from issues related to
internal and external validity.  To mitigate threats to internal validity, we
used the implementations of the testing tools provided by their authors (the tool
versions used are listed in Table~\ref{tool_overview_table}). 
Our implementation consists
of Python code for analyzing the log files of the web services to compute unique
500 errors, failure points, and library failure points. We thoroughly tested and spot checked the code and manually checked the testing results (coverage and faults) for a sample of the tools and web
services to gain confidence in the validity of our results.

As for the threats to external validity, because our evaluation is based on a
benchmark of 20 RESTful services, the results may not generalize. We note,
nevertheless, that our benchmark includes a diverse set of services,
including services that have been used in prior evaluations of REST API testing techniques.

Another potential threat is that we ran the services locally, which may result in different behavior than a real-world deployment.  For example, we set up a private Ethereum network instead of using the Ethereum main network for the ERC-20 service. However, we believe that this is unlikely to affect our results in any significant way. Furthermore, we manually checked that the locally-installed services behaved as expected, including for the ERC-20 deployment.

In our empirical study, we used the OpenAPI specifications provided with each web service in the benchmark.  We noticed that, for some services, the specification is incomplete, which may limit the ability of the tools to achieve higher coverage or finding more bugs. 
To investigate the prevalence of this phenomenon, we investigated over 1,000 specifications available on \url{https://apis.guru/} and checked them with Schemathesis; we found that a vast majority (almost 99\%) of the specifications contain some mismatch. In other words, it seems to be a common situation to have imperfect specifications, so the ones we used are representative of what the tools would have to rely upon in real-world scenarios.

\commentt{
\myeongsoo{To keep our space limit (500GB), we made a script that checks and deletes log files if the tools do not leverage them. To check the tools that use log files while fuzzing, we studied the tools' README files and papers. However, it is possible that the testing tools use the log files without specifying it in their paper or README file. We made sure that we did not delete the log file if it harmed the tools' performance with one-hour test runs for each RESTful service and testing tool. For 24 hours tests, we did not check if the log files that we deleted affected the tools' performance. However, we believe that the one-hour test runs would be enough to prevent it.}
\saurabh{Not sure that this text needs to be added}
\alex{I would not add it. It is very confusing and can open a can of worms}
}

RESTest requires additional information over what is in OpenAPI specifications. Specifically, it requires inter-parameter dependencies information~\cite{martin2019catalogue}. We therefore studied the parameters of each API endpoint for the web services in our benchmarks and added such dependencies (if any) as annotations to the OpenAPI specifications. Although it was simple to create such annotations, and it typically took us only a few minutes per specification, the quality of the specifications might have affected the performance of the tool. Unfortunately, this is an unavoidable threat when using RESTest.

Finally, EvoMasterWB requires, for each service under test, a test driver that performs various operations (e.g., starting and stopping the web service). Building such driver can require a non-trivial amount of manual effort. For the 10 web services we selected from previous work, we obtained the existing drivers created by the EvoMasterWB author. For the remaining 10 web services, however, we had to create a driver ourselves. Also in this case, this was a fairly trivial task, which we were able to complete in just a few minutes for each web service. And also in this case, the only way to avoid this threat would be to exclude EvoMasterWB from the set of tools considered, which would considerably decrease the value of our comparative study.

\vspace{-8pt}
\section{Related Work}
\label{sec:related}

Several recent studies have compared the effectiveness of testing techniques for REST services. Concurrent to ours, in particular, are three studies performed by Corradini et al.~\cite{corradini2021empirical}, Martin-Lopez, Segura, and Ruiz-Cort\'{e}s~\cite{martinlopez2021comparing}, and by Hatfield-Dodds and Dygalo~\cite{Zac2021schemathesis}.

Corradini et al.~\cite{corradini2021empirical} compared four black-box testing tools in terms of tool robustness and API coverage~\cite{corradini2021restats}. Our study includes all the tools they used and six additional tools. The code coverage metrics used in our study directly measure the amount of service code exercised and are different from the API coverage criteria~\cite{martinlopez2019criteria} computed by the Restats tool~\cite{corradini2021restats} and used in their study. For example, the parameter value metric can only be applied to Boolean or enumeration types. Moreover, our study performs a more thorough assessment of the techniques' bug-finding ability by counting the number of unique 5xx errors, failure points, and library failure points instead of simply checking whether a 5xx status code is returned, as was done in their study. Finally, we provide a detailed discussion of the strengths and weaknesses of the testing strategies employed by the tools and the implications of our findings for future technique and tool development in this area.

Martin-Lopez, Segura, and Ruiz-Cort\'{e}s compared EvoMasterWB, RESTest, and a hybrid of the two on a set of four REST APIs. Our study is different in that we compared EvoMasterWB with a set of 9 black-box tools (including RESTest) on a benchmark of 20 APIs. They showed that RESTest achieves better coverage than EvoMasterWB and that their bug-finding abilities are comparable. This is different from our result, which indicates that EvoMasterWB outperforms RESTest in terms of both coverage achieved and number of failures triggered. We believe this is partially because their study was based on only four services, which are all from the RESTest's benchmark dataset. Also, the manual work performed for tool configuration could have affected the results. In our study, we tried to avoid this issue by following the configuration method provided in the tools' papers and manuals~\cite{artifact}. We also collected tools and services in a systematic way, as explained in Sections~\ref{sec:tools} and~\ref{sec:benchmark}. Compared to their study, ours also contains a more detailed analysis of code coverage achieved and 500 errors found, as well as an additional discussion of the implications of our results.

In their study, Hatfield-Dodds and Dygalo~\cite{Zac2021schemathesis} compared Schemathesis to other black-box tools. Similar to Corradini et al.'s work, their work does not contain an in-depth analysis of the code coverage achieved and of the errors found by the tools beyond counting the number of errors triggered. Rather than providing a comparison of existing tools to assess their effectiveness, their work focused on the evaluation of Schemathesis.

In Section~\ref{sec:tools}, we provided a summary of 10 state-of-the-art techniques and tools that perform automated testing of REST APIs and that we considered in our evaluation.
We also discussed why we did not include in our study some of the tools we found. We note that there are other techniques that also target REST API testing. 
Among these, the approach proposed by Segura et al.~\cite{segura2017metamorphic} leverages metamorphic relationships to produce test cases.
We do not include this work in our study because their testing approach
is not automated: it requires a tester to provide low-level testing details,
including identifying metamorphic patterns and identifying problem-specific
metamorphic relations, which is non-trivial.
Godefroid, Lehmann, and Polishchuk~\cite{godefroid2020differential} introduced an approach
that performs differential testing to find regressions in an evolving specification and in the corresponding service.
Chakrabarti and Rodriquez~\cite{10.1145/1730874.1730902}
developed an approach that tests the connectedness (i.e., resource reachability) of RESTful services.
The approaches proposed by Chakrabarti and Kumar~\cite{chakrabarti2009test} and
by Reza and Gilst~\cite{reza2010framework}
rely on an XML-format specification for REST API testing.
The approaches developed by Seijas, Li, and Thompson~\cite{lamela2013towards}, 
by Fertig and Braun~\cite{fertig2015model},
and by Pinheiro, Endo, and Simao~\cite{pinheiro2013model} perform model-based testing. 
We did not include these techniques in our study because
they use a tool-specific REST API specification~\cite{chakrabarti2009test,reza2010framework} that requires non-trivial information to be provided by the user, rely on formal specifications~\cite{lamela2013towards,fertig2015model,pinheiro2013model}, 
or target specific problems~\cite{godefroid2020differential,10.1145/1730874.1730902}.

Recently, after this work was finalized, two additional techniques and tools were proposed that perform combinatorial testing~\cite{wu2022combinatorial} and model-based testing~\cite{liu2022morest} of RESTful APIs. Although we could not consider them in our current comparison, we will evaluate these tools for inclusion in an extended version of this work.

Numerous tools and libraries provide support for
REST API testing, such as Postman~\cite{postman}, REST Assured~\cite{restassured}, 
ReadyAPI~\cite{readyapi}, and API Fortress~\cite{apifortress}. 
We did not include them in our study because they do not support
automated testing. There are also tools that
focus on tracking, fuzzing, and replaying API traffic such as 
AppSpider~\cite{appspider} and WAS~\cite{was}.
We did not include these tools as they target fuzzing and require
pre-recorded API usage information.

Finally, there exist many techniques and tools designed to
test SOAP web services with WSDL-based specifications (e.g.,~\cite{bai2005wsdl, bartolini2009ws, ma2008wsdl, sneed2006wsdltest, tsai2002coyote})
and more broadly target the testing of service-oriented architectures (e.g.,~\cite{canfora2007service, bozkurt2013testing}).
We do not include these techniques and tools in our study, as we focus on REST API testing.

\vspace{-4pt}
\section{Conclusion and Future Work}
\label{sec:conclusion}

To gain insights into the effectiveness of existing REST API testing techniques and tools, we performed an empirical study in which we applied 10 state-of-the-art techniques to 20 RESTful web services and compared them in terms of code coverage achieved and unique failures triggered. We presented the results of the study, along with an analysis of the strengths and weaknesses of the techniques, summarized the lessons learned, and discussed implications for future research. Our experiment infrastructure, data, and results are publicly available~\cite{artifact}.
In future work, we will extend our evaluation by using mutation to further evaluate the tools' fault-detection ability. 
We will also leverage the insights gained from our study and preliminary investigations to develop new techniques for testing REST APIs that perform better input parameter generation and consider dependencies among operations. Specifically, we will investigate ways to extract meaningful input values from the API specification and server logs, study the application of symbolic analysis to extract relevant information from the code, and research the use of NLP-based techniques to infer producer-consumer dependencies between operations.

\vspace{-4pt}
\section*{Data-Availability Statement}

Data and code for reproducing our results are available on Zenodo~\cite{artifact-issta22}.
Updated information about the project can be found at \url{https://bit.ly/RESTTestToolsStudy}.

\vspace{-4pt}
\section*{Acknowledgments}
\vspace{-4pt}
  This work was partially supported by
  NSF, under grant CCF-0725202,
  DARPA, under contract N66001-21-C-4024,
  DOE, under contract DE-FOA-0002460,
  and gifts from Facebook, Google, IBM Research, and Microsoft Research.

\balance

\bibliographystyle{ACM-Reference-Format}
\bibliography{paper}


\begin{thebibliography}{87}


\ifx \showCODEN    \undefined \def \showCODEN     #1{\unskip}     \fi
\ifx \showDOI      \undefined \def \showDOI       #1{#1}\fi
\ifx \showISBNx    \undefined \def \showISBNx     #1{\unskip}     \fi
\ifx \showISBNxiii \undefined \def \showISBNxiii  #1{\unskip}     \fi
\ifx \showISSN     \undefined \def \showISSN      #1{\unskip}     \fi
\ifx \showLCCN     \undefined \def \showLCCN      #1{\unskip}     \fi
\ifx \shownote     \undefined \def \shownote      #1{#1}          \fi
\ifx \showarticletitle \undefined \def \showarticletitle #1{#1}   \fi
\ifx \showURL      \undefined \def \showURL       {\relax}        \fi
\providecommand\bibfield[2]{#2}
\providecommand\bibinfo[2]{#2}
\providecommand\natexlab[1]{#1}
\providecommand\showeprint[2][]{arXiv:#2}

\bibitem[APIBlueprint(2021)]%
        {apiblueprint}
APIBlueprint \bibinfo{year}{2021}\natexlab{}.
\newblock \bibinfo{title}{API Blueprint}.
\newblock
\newblock
\urldef\tempurl%
\url{https://apiblueprint.org/}
\showURL{%
\tempurl}
\newblock
\shownote{Accessed: Jun 3, 2022}.


\bibitem[APIFortress(2022)]%
        {apifortress}
APIFortress \bibinfo{year}{2022}\natexlab{}.
\newblock \bibinfo{title}{API Fortress}.
\newblock
\newblock
\urldef\tempurl%
\url{https://apifortress.com}
\showURL{%
\tempurl}
\newblock
\shownote{Accessed: Jun 3, 2022}.


\bibitem[APIFuzzer(2022)]%
        {apifuzzer}
APIFuzzer \bibinfo{year}{2022}\natexlab{}.
\newblock \bibinfo{title}{APIFuzzer}.
\newblock
\newblock
\urldef\tempurl%
\url{https://github.com/KissPeter/APIFuzzer}
\showURL{%
\tempurl}
\newblock
\shownote{Accessed: Jun 3, 2022}.


\bibitem[apisguru(2022)]%
        {apisguru}
apisguru \bibinfo{year}{2022}\natexlab{}.
\newblock \bibinfo{title}{APIs.guru API Directory}.
\newblock
\newblock
\urldef\tempurl%
\url{https://apis.guru/}
\showURL{%
\tempurl}
\newblock
\shownote{Accessed: Jun 3, 2022}.


\bibitem[AppSpider(2022)]%
        {appspider}
AppSpider \bibinfo{year}{2022}\natexlab{}.
\newblock \bibinfo{title}{AppSpider}.
\newblock
\newblock
\urldef\tempurl%
\url{https://www.rapid7.com/products/appspider}
\showURL{%
\tempurl}
\newblock
\shownote{Accessed: Jun 3, 2022}.


\bibitem[Arcuri(2017)]%
        {arcuri2017restful}
\bibfield{author}{\bibinfo{person}{Andrea Arcuri}.}
  \bibinfo{year}{2017}\natexlab{}.
\newblock \showarticletitle{RESTful API automated test case generation}. In
  \bibinfo{booktitle}{\emph{2017 IEEE International Conference on Software
  Quality, Reliability and Security (QRS)}}. \bibinfo{publisher}{IEEE},
  \bibinfo{address}{Prague, Czech Republic}, \bibinfo{pages}{9--20}.
\newblock


\bibitem[Arcuri(2018)]%
        {arcuri2018evomaster}
\bibfield{author}{\bibinfo{person}{Andrea Arcuri}.}
  \bibinfo{year}{2018}\natexlab{}.
\newblock \showarticletitle{Evomaster: Evolutionary multi-context automated
  system test generation}. In \bibinfo{booktitle}{\emph{2018 IEEE 11th
  International Conference on Software Testing, Verification and Validation
  (ICST)}}. \bibinfo{publisher}{IEEE}, \bibinfo{address}{Västerås, Sweden},
  \bibinfo{pages}{394--397}.
\newblock


\bibitem[Arcuri(2019a)]%
        {DBLP:journals/corr/abs-1901-01541}
\bibfield{author}{\bibinfo{person}{Andrea Arcuri}.}
  \bibinfo{year}{2019}\natexlab{a}.
\newblock \showarticletitle{Many Independent Objective {(MIO)} Algorithm for
  Test Suite Generation}.
\newblock \bibinfo{journal}{\emph{CoRR}}  \bibinfo{volume}{abs/1901.01541}
  (\bibinfo{year}{2019}), \bibinfo{pages}{3--17}.
\newblock
\showeprint[arXiv]{1901.01541}
\urldef\tempurl%
\url{http://arxiv.org/abs/1901.01541}
\showURL{%
\tempurl}


\bibitem[Arcuri(2019b)]%
        {arcuri2019restful}
\bibfield{author}{\bibinfo{person}{Andrea Arcuri}.}
  \bibinfo{year}{2019}\natexlab{b}.
\newblock \showarticletitle{RESTful API automated test case generation with
  EvoMaster}.
\newblock \bibinfo{journal}{\emph{ACM Transactions on Software Engineering and
  Methodology (TOSEM)}} \bibinfo{volume}{28}, \bibinfo{number}{1}
  (\bibinfo{year}{2019}), \bibinfo{pages}{1--37}.
\newblock


\bibitem[Arcuri(2020)]%
        {arcuri2020automated}
\bibfield{author}{\bibinfo{person}{Andrea Arcuri}.}
  \bibinfo{year}{2020}\natexlab{}.
\newblock \showarticletitle{Automated Black-and White-Box Testing of RESTful
  APIs With EvoMaster}.
\newblock \bibinfo{journal}{\emph{IEEE Software}} \bibinfo{volume}{38},
  \bibinfo{number}{3} (\bibinfo{year}{2020}), \bibinfo{pages}{72--78}.
\newblock


\bibitem[Arcuri and Galeotti(2019)]%
        {DBLP:conf/gecco/ArcuriG19}
\bibfield{author}{\bibinfo{person}{Andrea Arcuri} {and}
  \bibinfo{person}{Juan~P. Galeotti}.} \bibinfo{year}{2019}\natexlab{}.
\newblock \showarticletitle{{SQL} data generation to enhance search-based
  system testing}. In \bibinfo{booktitle}{\emph{Proceedings of the Genetic and
  Evolutionary Computation Conference, {GECCO} July 13-17, 2019}},
  \bibfield{editor}{\bibinfo{person}{Anne Auger} {and} \bibinfo{person}{Thomas
  St{\"{u}}tzle}} (Eds.). \bibinfo{publisher}{{ACM}}, \bibinfo{address}{Prague,
  Czech Republic}, \bibinfo{pages}{1390--1398}.
\newblock
\urldef\tempurl%
\url{https://doi.org/10.1145/3321707.3321732}
\showDOI{\tempurl}


\bibitem[Artifact(2022)]%
        {artifact}
Artifact \bibinfo{year}{2022}\natexlab{}.
\newblock \bibinfo{title}{Companion page with experiment infrastructure, data,
  and results}.
\newblock
\newblock
\urldef\tempurl%
\url{bit.ly/RESTTestToolsStudy}
\showURL{%
\tempurl}
\newblock
\shownote{Accessed: Jun 3, 2022}.


\bibitem[Atlidakis et~al\mbox{.}(2020a)]%
        {atlidakis2020pythia}
\bibfield{author}{\bibinfo{person}{Vaggelis Atlidakis}, \bibinfo{person}{Roxana
  Geambasu}, \bibinfo{person}{Patrice Godefroid}, \bibinfo{person}{Marina
  Polishchuk}, {and} \bibinfo{person}{Baishakhi Ray}.}
  \bibinfo{year}{2020}\natexlab{a}.
\newblock \bibinfo{title}{Pythia: Grammar-Based Fuzzing of REST APIs with
  Coverage-guided Feedback and Learning-based Mutations}.
\newblock
\newblock
\showeprint[arxiv]{2005.11498}~[cs.SE]


\bibitem[Atlidakis et~al\mbox{.}(2019)]%
        {atlidakis2019restler}
\bibfield{author}{\bibinfo{person}{Vaggelis Atlidakis},
  \bibinfo{person}{Patrice Godefroid}, {and} \bibinfo{person}{Marina
  Polishchuk}.} \bibinfo{year}{2019}\natexlab{}.
\newblock \showarticletitle{Restler: Stateful rest api fuzzing}. In
  \bibinfo{booktitle}{\emph{2019 IEEE/ACM 41st International Conference on
  Software Engineering (ICSE)}}. \bibinfo{publisher}{IEEE},
  \bibinfo{address}{Montreal, QC, Canada}, \bibinfo{pages}{748--758}.
\newblock


\bibitem[Atlidakis et~al\mbox{.}(2020b)]%
        {atlidakis2020checking}
\bibfield{author}{\bibinfo{person}{Vaggelis Atlidakis},
  \bibinfo{person}{Patrice Godefroid}, {and} \bibinfo{person}{Marina
  Polishchuk}.} \bibinfo{year}{2020}\natexlab{b}.
\newblock \showarticletitle{Checking Security Properties of Cloud Service REST
  APIs}. In \bibinfo{booktitle}{\emph{13th International Conference on Software
  Testing, Validation and Verification (ICST)}}. \bibinfo{publisher}{IEEE},
  \bibinfo{address}{Porto, Portugal}, \bibinfo{pages}{387--397}.
\newblock


\bibitem[Bai et~al\mbox{.}(2005)]%
        {bai2005wsdl}
\bibfield{author}{\bibinfo{person}{Xiaoying Bai}, \bibinfo{person}{Wenli Dong},
  \bibinfo{person}{Wei-Tek Tsai}, {and} \bibinfo{person}{Yinong Chen}.}
  \bibinfo{year}{2005}\natexlab{}.
\newblock \showarticletitle{WSDL-based automatic test case generation for web
  services testing}. In \bibinfo{booktitle}{\emph{IEEE International Workshop
  on Service-Oriented System Engineering (SOSE)}}. \bibinfo{publisher}{IEEE},
  \bibinfo{address}{Beijing, China}, \bibinfo{pages}{207--212}.
\newblock


\bibitem[Baldoni et~al\mbox{.}(2018)]%
        {baldoni2018survey}
\bibfield{author}{\bibinfo{person}{Roberto Baldoni}, \bibinfo{person}{Emilio
  Coppa}, \bibinfo{person}{Daniele~Cono D’elia}, \bibinfo{person}{Camil
  Demetrescu}, {and} \bibinfo{person}{Irene Finocchi}.}
  \bibinfo{year}{2018}\natexlab{}.
\newblock \showarticletitle{A survey of symbolic execution techniques}.
\newblock \bibinfo{journal}{\emph{ACM Computing Surveys (CSUR)}}
  \bibinfo{volume}{51}, \bibinfo{number}{3} (\bibinfo{year}{2018}),
  \bibinfo{pages}{1--39}.
\newblock


\bibitem[Bartolini et~al\mbox{.}(2009)]%
        {bartolini2009ws}
\bibfield{author}{\bibinfo{person}{Cesare Bartolini}, \bibinfo{person}{Antonia
  Bertolino}, \bibinfo{person}{Eda Marchetti}, {and} \bibinfo{person}{Andrea
  Polini}.} \bibinfo{year}{2009}\natexlab{}.
\newblock \showarticletitle{WS-TAXI: A WSDL-based testing tool for web
  services}. In \bibinfo{booktitle}{\emph{International Conference on Software
  Testing Verification and Validation (ICST)}}. \bibinfo{publisher}{IEEE},
  \bibinfo{address}{Denver, CO, USA}, \bibinfo{pages}{326--335}.
\newblock


\bibitem[bBOXRT(2022)]%
        {bboxrttool}
bBOXRT \bibinfo{year}{2022}\natexlab{}.
\newblock \bibinfo{title}{bBOXRT}.
\newblock
\newblock
\urldef\tempurl%
\url{https://git.dei.uc.pt/cnl/bBOXRT}
\showURL{%
\tempurl}
\newblock
\shownote{Accessed: Jun 3, 2022}.


\bibitem[Bozkurt et~al\mbox{.}(2013)]%
        {bozkurt2013testing}
\bibfield{author}{\bibinfo{person}{Mustafa Bozkurt}, \bibinfo{person}{Mark
  Harman}, {and} \bibinfo{person}{Youssef Hassoun}.}
  \bibinfo{year}{2013}\natexlab{}.
\newblock \showarticletitle{Testing and verification in service-oriented
  architecture: a survey}.
\newblock \bibinfo{journal}{\emph{Software Testing, Verification and
  Reliability}} \bibinfo{volume}{23}, \bibinfo{number}{4}
  (\bibinfo{year}{2013}), \bibinfo{pages}{261--313}.
\newblock


\bibitem[Canfora and Di~Penta(2007)]%
        {canfora2007service}
\bibfield{author}{\bibinfo{person}{Gerardo Canfora} {and}
  \bibinfo{person}{Massimiliano Di~Penta}.} \bibinfo{year}{2007}\natexlab{}.
\newblock \showarticletitle{Service-oriented architectures testing: A survey}.
\newblock In \bibinfo{booktitle}{\emph{Software Engineering}}.
  \bibinfo{publisher}{Springer}, \bibinfo{address}{Berlin, Heidelberg},
  \bibinfo{pages}{78--105}.
\newblock


\bibitem[Cats(2022)]%
        {cats}
Cats \bibinfo{year}{2022}\natexlab{}.
\newblock \bibinfo{title}{Cats}.
\newblock
\newblock
\urldef\tempurl%
\url{https://github.com/Endava/cats}
\showURL{%
\tempurl}
\newblock
\shownote{Accessed: Jun 3, 2022}.


\bibitem[Chakrabarti and Kumar(2009)]%
        {chakrabarti2009test}
\bibfield{author}{\bibinfo{person}{Sujit~Kumar Chakrabarti} {and}
  \bibinfo{person}{Prashant Kumar}.} \bibinfo{year}{2009}\natexlab{}.
\newblock \showarticletitle{Test-the-rest: An approach to testing restful
  web-services}. In \bibinfo{booktitle}{\emph{2009 Computation World: Future
  Computing, Service Computation, Cognitive, Adaptive, Content, Patterns}}.
  \bibinfo{publisher}{IEEE}, \bibinfo{address}{Athens, Greece},
  \bibinfo{pages}{302--308}.
\newblock


\bibitem[Chakrabarti and Rodriquez(2010)]%
        {10.1145/1730874.1730902}
\bibfield{author}{\bibinfo{person}{Sujit~Kumar Chakrabarti} {and}
  \bibinfo{person}{Reswin Rodriquez}.} \bibinfo{year}{2010}\natexlab{}.
\newblock \showarticletitle{Connectedness Testing of RESTful Web-Services}. In
  \bibinfo{booktitle}{\emph{Proceedings of the 3rd India Software Engineering
  Conference}} (Mysore, India) \emph{(\bibinfo{series}{ISEC '10})}.
  \bibinfo{publisher}{Association for Computing Machinery},
  \bibinfo{address}{New York, NY, USA}, \bibinfo{pages}{143–152}.
\newblock
\showISBNx{9781605589220}
\urldef\tempurl%
\url{https://doi.org/10.1145/1730874.1730902}
\showDOI{\tempurl}


\bibitem[Corradini et~al\mbox{.}(2021a)]%
        {corradini2021empirical}
\bibfield{author}{\bibinfo{person}{Davide Corradini}, \bibinfo{person}{Amedeo
  Zampieri}, \bibinfo{person}{Michele Pasqua}, {and} \bibinfo{person}{Mariano
  Ceccato}.} \bibinfo{year}{2021}\natexlab{a}.
\newblock \showarticletitle{Empirical comparison of black-box test case
  generation tools for RESTful APIs}. In \bibinfo{booktitle}{\emph{2021 IEEE
  21st International Working Conference on Source Code Analysis and
  Manipulation (SCAM)}}. \bibinfo{publisher}{IEEE},
  \bibinfo{address}{Luxembourg}, \bibinfo{pages}{226--236}.
\newblock


\bibitem[Corradini et~al\mbox{.}(2021b)]%
        {corradini2021restats}
\bibfield{author}{\bibinfo{person}{Davide Corradini}, \bibinfo{person}{Amedeo
  Zampieri}, \bibinfo{person}{Michele Pasqua}, {and} \bibinfo{person}{Mariano
  Ceccato}.} \bibinfo{year}{2021}\natexlab{b}.
\newblock \showarticletitle{Restats: A test coverage tool for RESTful APIs}. In
  \bibinfo{booktitle}{\emph{2021 IEEE International Conference on Software
  Maintenance and Evolution (ICSME)}}. \bibinfo{publisher}{IEEE},
  \bibinfo{address}{Luxembourg}, \bibinfo{pages}{594--598}.
\newblock


\bibitem[Corradini et~al\mbox{.}(2022)]%
        {corradiniautomated}
\bibfield{author}{\bibinfo{person}{Davide Corradini}, \bibinfo{person}{Amedeo
  Zampieri}, \bibinfo{person}{Michele Pasqua}, \bibinfo{person}{Emanuele
  Viglianisi}, \bibinfo{person}{Michael Dallago}, {and}
  \bibinfo{person}{Mariano Ceccato}.} \bibinfo{year}{2022}\natexlab{}.
\newblock \showarticletitle{Automated black-box testing of nominal and error
  scenarios in RESTful APIs}.
\newblock \bibinfo{journal}{\emph{Software Testing, Verification and
  Reliability}} (\bibinfo{year}{2022}), \bibinfo{pages}{e1808}.
\newblock


\bibitem[Dredd(2022)]%
        {dredd}
Dredd \bibinfo{year}{2022}\natexlab{}.
\newblock \bibinfo{title}{Dredd}.
\newblock
\newblock
\urldef\tempurl%
\url{https://github.com/apiaryio/dredd}
\showURL{%
\tempurl}
\newblock
\shownote{Accessed: may 1, 2022}.


\bibitem[Ed-Douibi et~al\mbox{.}(2018)]%
        {ed2018automatic}
\bibfield{author}{\bibinfo{person}{Hamza Ed-Douibi}, \bibinfo{person}{Javier
  Luis~C{\'a}novas Izquierdo}, {and} \bibinfo{person}{Jordi Cabot}.}
  \bibinfo{year}{2018}\natexlab{}.
\newblock \showarticletitle{Automatic generation of test cases for REST APIs: a
  specification-based approach}. In \bibinfo{booktitle}{\emph{22nd
  International Enterprise Distributed Object Computing Conference (EDOC)}}.
  IEEE, \bibinfo{pages}{181--190}.
\newblock


\bibitem[EvoMaster(2022)]%
        {evomastertool}
EvoMaster \bibinfo{year}{2022}\natexlab{}.
\newblock \bibinfo{title}{EvoMaster}.
\newblock
\newblock
\urldef\tempurl%
\url{https://github.com/EMResearch/EvoMaster}
\showURL{%
\tempurl}
\newblock
\shownote{Accessed: Jun 3, 2022}.


\bibitem[Fertig and Braun(2015)]%
        {fertig2015model}
\bibfield{author}{\bibinfo{person}{Tobias Fertig} {and} \bibinfo{person}{Peter
  Braun}.} \bibinfo{year}{2015}\natexlab{}.
\newblock \showarticletitle{Model-driven testing of restful apis}. In
  \bibinfo{booktitle}{\emph{Proceedings of the 24th International Conference on
  World Wide Web}}. \bibinfo{pages}{1497--1502}.
\newblock


\bibitem[Fielding(2000)]%
        {fielding2000architectural}
\bibfield{author}{\bibinfo{person}{Roy~T Fielding}.}
  \bibinfo{year}{2000}\natexlab{}.
\newblock \bibinfo{booktitle}{\emph{Architectural styles and the design of
  network-based software architectures}}. Vol.~\bibinfo{volume}{7}.
\newblock \bibinfo{publisher}{University of California, Irvine Irvine}.
\newblock


\bibitem[Freedman et~al\mbox{.}(2007)]%
        {freedman2007statistics}
\bibfield{author}{\bibinfo{person}{David Freedman}, \bibinfo{person}{Robert
  Pisani}, {and} \bibinfo{person}{Roger Purves}.}
  \bibinfo{year}{2007}\natexlab{}.
\newblock \bibinfo{booktitle}{\emph{Statistics (international student
  edition)}}.
\newblock \bibinfo{publisher}{WW Norton \& Company}.
\newblock


\bibitem[Gavel(2022)]%
        {gavel}
Gavel \bibinfo{year}{2022}\natexlab{}.
\newblock \bibinfo{title}{Gavel}.
\newblock
\newblock
\urldef\tempurl%
\url{https://github.com/apiaryio/gavel.js}
\showURL{%
\tempurl}
\newblock
\shownote{Accessed: Jun 3, 2022}.


\bibitem[Godefroid et~al\mbox{.}(2020a)]%
        {godefroid2020intelligent}
\bibfield{author}{\bibinfo{person}{Patrice Godefroid}, \bibinfo{person}{Bo-Yuan
  Huang}, {and} \bibinfo{person}{Marina Polishchuk}.}
  \bibinfo{year}{2020}\natexlab{a}.
\newblock \showarticletitle{Intelligent REST API data fuzzing}. In
  \bibinfo{booktitle}{\emph{Proceedings of the 28th ACM Joint Meeting on
  European Software Engineering Conference and Symposium on the Foundations of
  Software Engineering}}. \bibinfo{pages}{725--736}.
\newblock


\bibitem[Godefroid et~al\mbox{.}(2020b)]%
        {godefroid2020differential}
\bibfield{author}{\bibinfo{person}{Patrice Godefroid}, \bibinfo{person}{Daniel
  Lehmann}, {and} \bibinfo{person}{Marina Polishchuk}.}
  \bibinfo{year}{2020}\natexlab{b}.
\newblock \showarticletitle{Differential regression testing for REST APIs}. In
  \bibinfo{booktitle}{\emph{Proceedings of the 29th ACM SIGSOFT International
  Symposium on Software Testing and Analysis (ISSTA)}}.
  \bibinfo{pages}{312--323}.
\newblock


\bibitem[GotSwag(2018)]%
        {gotswag}
GotSwag \bibinfo{year}{2018}\natexlab{}.
\newblock \bibinfo{title}{GotSwag}.
\newblock
\newblock
\urldef\tempurl%
\url{https://github.com/mobilcom-debitel/got-swag}
\showURL{%
\tempurl}
\newblock
\shownote{Accessed: Jun 3, 2022}.


\bibitem[Hatfield-Dodds and Dygalo(2021)]%
        {Zac2021schemathesis}
\bibfield{author}{\bibinfo{person}{Zac Hatfield-Dodds} {and}
  \bibinfo{person}{Dmitry Dygalo}.} \bibinfo{year}{2021}\natexlab{}.
\newblock \showarticletitle{Deriving Semantics-Aware Fuzzers from Web API
  Schemas}.
\newblock \bibinfo{journal}{\emph{arXiv preprint arXiv:2112.10328}}
  (\bibinfo{year}{2021}).
\newblock


\bibitem[Hypothesis(2022)]%
        {hypothesis}
Hypothesis \bibinfo{year}{2022}\natexlab{}.
\newblock \bibinfo{title}{Hypothesis}.
\newblock
\newblock
\urldef\tempurl%
\url{https://hypothesis.works/}
\showURL{%
\tempurl}
\newblock
\shownote{Accessed: Jun 3, 2022}.


\bibitem[IDLReasoner(2022)]%
        {idlreasoner}
IDLReasoner \bibinfo{year}{2022}\natexlab{}.
\newblock \bibinfo{title}{IDLReasoner}.
\newblock
\newblock
\urldef\tempurl%
\url{https://github.com/isa-group/IDLReasoner}
\showURL{%
\tempurl}
\newblock
\shownote{Accessed: May 1, 2022}.


\bibitem[JaCoCo(2021)]%
        {jacoco}
JaCoCo \bibinfo{year}{2021}\natexlab{}.
\newblock \bibinfo{title}{JaCoCo}.
\newblock
\newblock
\urldef\tempurl%
\url{https://www.eclemma.org/jacoco/}
\showURL{%
\tempurl}
\newblock
\shownote{Accessed: Jun 3, 2022}.


\bibitem[Jordan and Mitchell(2015)]%
        {jordan2015machine}
\bibfield{author}{\bibinfo{person}{Michael~I Jordan} {and}
  \bibinfo{person}{Tom~M Mitchell}.} \bibinfo{year}{2015}\natexlab{}.
\newblock \showarticletitle{Machine learning: Trends, perspectives, and
  prospects}.
\newblock \bibinfo{journal}{\emph{Science}} (\bibinfo{year}{2015}),
  \bibinfo{pages}{255--260}.
\newblock


\bibitem[Karlsson et~al\mbox{.}(2020a)]%
        {karlsson2020automatic}
\bibfield{author}{\bibinfo{person}{Stefan Karlsson}, \bibinfo{person}{Adnan
  {\v{C}}au{\v{s}}evi{\'c}}, {and} \bibinfo{person}{Daniel Sundmark}.}
  \bibinfo{year}{2020}\natexlab{a}.
\newblock \showarticletitle{Automatic Property-based Testing of GraphQL APIs}.
\newblock \bibinfo{journal}{\emph{arXiv preprint arXiv:2012.07380}}
  (\bibinfo{year}{2020}).
\newblock


\bibitem[Karlsson et~al\mbox{.}(2020b)]%
        {karlsson2020quickrest}
\bibfield{author}{\bibinfo{person}{Stefan Karlsson}, \bibinfo{person}{Adnan
  {\v{C}}au{\v{s}}evi{\'c}}, {and} \bibinfo{person}{Daniel Sundmark}.}
  \bibinfo{year}{2020}\natexlab{b}.
\newblock \showarticletitle{QuickREST: Property-based Test Generation of
  OpenAPI-Described RESTful APIs}. In \bibinfo{booktitle}{\emph{13th
  International Conference on Software Testing, Validation and Verification
  (ICST)}}. IEEE, \bibinfo{pages}{131--141}.
\newblock


\bibitem[Kim et~al\mbox{.}(2022)]%
        {artifact-issta22}
\bibfield{author}{\bibinfo{person}{Myeongsoo Kim}, \bibinfo{person}{Qi Xin},
  \bibinfo{person}{Saurabh Sinha}, {and} \bibinfo{person}{Alessandro Orso}.}
  \bibinfo{year}{2022}\natexlab{}.
\newblock \bibinfo{booktitle}{\emph{{Automated Test Generation for REST APIs:
  Replication Package}}}.
\newblock
\urldef\tempurl%
\url{https://doi.org/10.5281/zenodo.6534554}
\showDOI{\tempurl}


\bibitem[King(1976)]%
        {king1976symbolic}
\bibfield{author}{\bibinfo{person}{James~C King}.}
  \bibinfo{year}{1976}\natexlab{}.
\newblock \showarticletitle{Symbolic execution and program testing}.
\newblock \bibinfo{journal}{\emph{Commun. ACM}} \bibinfo{volume}{19},
  \bibinfo{number}{7} (\bibinfo{year}{1976}), \bibinfo{pages}{385--394}.
\newblock


\bibitem[Koopman et~al\mbox{.}(1997)]%
        {koopman1997comparing}
\bibfield{author}{\bibinfo{person}{Philip Koopman}, \bibinfo{person}{John
  Sung}, \bibinfo{person}{Christopher Dingman}, \bibinfo{person}{Daniel
  Siewiorek}, {and} \bibinfo{person}{Ted Marz}.}
  \bibinfo{year}{1997}\natexlab{}.
\newblock \showarticletitle{Comparing operating systems using robustness
  benchmarks}. In \bibinfo{booktitle}{\emph{Proceedings of 16th IEEE Symposium
  on Reliable Distributed Systems (SRDS)}}. IEEE, \bibinfo{pages}{72--79}.
\newblock


\bibitem[K{\"u}bler et~al\mbox{.}(2009)]%
        {kubler2009dependency}
\bibfield{author}{\bibinfo{person}{Sandra K{\"u}bler}, \bibinfo{person}{Ryan
  McDonald}, {and} \bibinfo{person}{Joakim Nivre}.}
  \bibinfo{year}{2009}\natexlab{}.
\newblock \showarticletitle{Dependency parsing}.
\newblock \bibinfo{journal}{\emph{Synthesis lectures on human language
  technologies}} (\bibinfo{year}{2009}), \bibinfo{pages}{1--127}.
\newblock


\bibitem[Kuhn et~al\mbox{.}(2013)]%
        {kuhn2013introduction}
\bibfield{author}{\bibinfo{person}{D~Richard Kuhn}, \bibinfo{person}{Raghu~N
  Kacker}, {and} \bibinfo{person}{Yu Lei}.} \bibinfo{year}{2013}\natexlab{}.
\newblock \bibinfo{booktitle}{\emph{Introduction to combinatorial testing}}.
\newblock \bibinfo{publisher}{CRC press}.
\newblock


\bibitem[Lamela~Seijas et~al\mbox{.}(2013)]%
        {lamela2013towards}
\bibfield{author}{\bibinfo{person}{Pablo Lamela~Seijas},
  \bibinfo{person}{Huiqing Li}, {and} \bibinfo{person}{Simon Thompson}.}
  \bibinfo{year}{2013}\natexlab{}.
\newblock \showarticletitle{Towards property-based testing of RESTful web
  services}. In \bibinfo{booktitle}{\emph{Proceedings of the twelfth ACM
  SIGPLAN workshop on Erlang}}. \bibinfo{pages}{77--78}.
\newblock


\bibitem[Laranjeiro et~al\mbox{.}(2021)]%
        {laranjeiro2021black}
\bibfield{author}{\bibinfo{person}{Nuno Laranjeiro}, \bibinfo{person}{Jo{\~a}o
  Agnelo}, {and} \bibinfo{person}{Jorge Bernardino}.}
  \bibinfo{year}{2021}\natexlab{}.
\newblock \showarticletitle{A Black Box Tool for Robustness Testing of REST
  Services}.
\newblock \bibinfo{journal}{\emph{IEEE Access}} (\bibinfo{year}{2021}),
  \bibinfo{pages}{24738--24754}.
\newblock


\bibitem[Liu et~al\mbox{.}(2022)]%
        {liu2022morest}
\bibfield{author}{\bibinfo{person}{Yi Liu}, \bibinfo{person}{Yuekang Li},
  \bibinfo{person}{Gelei Deng}, \bibinfo{person}{Yang Liu},
  \bibinfo{person}{Ruiyuan Wan}, \bibinfo{person}{Runchao Wu},
  \bibinfo{person}{Dandan Ji}, \bibinfo{person}{Shiheng Xu}, {and}
  \bibinfo{person}{Minli Bao}.} \bibinfo{year}{2022}\natexlab{}.
\newblock \showarticletitle{Morest: Model-based RESTful API Testing with
  Execution Feedback}.
\newblock \bibinfo{journal}{\emph{arXiv preprint arXiv:2204.12148}}
  (\bibinfo{year}{2022}).
\newblock


\bibitem[Ma et~al\mbox{.}(2008)]%
        {ma2008wsdl}
\bibfield{author}{\bibinfo{person}{Chunyan Ma}, \bibinfo{person}{Chenglie Du},
  \bibinfo{person}{Tao Zhang}, \bibinfo{person}{Fei Hu}, {and}
  \bibinfo{person}{Xiaobin Cai}.} \bibinfo{year}{2008}\natexlab{}.
\newblock \showarticletitle{WSDL-based automated test data generation for web
  service}. In \bibinfo{booktitle}{\emph{2008 International Conference on
  Computer Science and Software Engineering}}, Vol.~\bibinfo{volume}{2}. IEEE,
  \bibinfo{pages}{731--737}.
\newblock


\bibitem[Manning and Schutze(1999)]%
        {manning1999foundations}
\bibfield{author}{\bibinfo{person}{Christopher Manning} {and}
  \bibinfo{person}{Hinrich Schutze}.} \bibinfo{year}{1999}\natexlab{}.
\newblock \bibinfo{booktitle}{\emph{Foundations of statistical natural language
  processing}}.
\newblock \bibinfo{publisher}{MIT press}.
\newblock


\bibitem[Martin-Lopez et~al\mbox{.}(2021b)]%
        {martinlopez2021}
\bibfield{author}{\bibinfo{person}{Alberto Martin-Lopez},
  \bibinfo{person}{Sergio Segura}, \bibinfo{person}{Carlos Muller}, {and}
  \bibinfo{person}{Antonio Ruiz-Cortes}.} \bibinfo{year}{2021}\natexlab{b}.
\newblock \showarticletitle{Specification and Automated Analysis of
  Inter-Parameter Dependencies in Web APIs}.
\newblock \bibinfo{journal}{\emph{IEEE Transactions on Services Computing}}
  (\bibinfo{year}{2021}), \bibinfo{pages}{1--1}.
\newblock
\urldef\tempurl%
\url{https://doi.org/10.1109/TSC.2021.3050610}
\showDOI{\tempurl}


\bibitem[Martin-Lopez et~al\mbox{.}(2019a)]%
        {martin2019catalogue}
\bibfield{author}{\bibinfo{person}{Alberto Martin-Lopez},
  \bibinfo{person}{Sergio Segura}, {and} \bibinfo{person}{Antonio
  Ruiz-Cort{\'e}s}.} \bibinfo{year}{2019}\natexlab{a}.
\newblock \showarticletitle{A catalogue of inter-parameter dependencies in
  RESTful web APIs}. In \bibinfo{booktitle}{\emph{International Conference on
  Service-Oriented Computing}}. Springer, \bibinfo{pages}{399--414}.
\newblock


\bibitem[Martin-Lopez et~al\mbox{.}(2019b)]%
        {martinlopez2019criteria}
\bibfield{author}{\bibinfo{person}{Alberto Martin-Lopez},
  \bibinfo{person}{Sergio Segura}, {and} \bibinfo{person}{Antonio
  Ruiz-Cort\'{e}s}.} \bibinfo{year}{2019}\natexlab{b}.
\newblock \showarticletitle{{Test Coverage Criteria for RESTful Web APIs}}. In
  \bibinfo{booktitle}{\emph{Proceedings of the 10th ACM SIGSOFT International
  Workshop on Automating TEST Case Design, Selection, and Evaluation}}.
  \bibinfo{pages}{15–21}.
\newblock


\bibitem[Martin-Lopez et~al\mbox{.}(2020)]%
        {martin2020restest}
\bibfield{author}{\bibinfo{person}{Alberto Martin-Lopez},
  \bibinfo{person}{Sergio Segura}, {and} \bibinfo{person}{Antonio
  Ruiz-Cort{\'e}s}.} \bibinfo{year}{2020}\natexlab{}.
\newblock \showarticletitle{RESTest: Black-Box Constraint-Based Testing of
  RESTful Web APIs}. In \bibinfo{booktitle}{\emph{International Conference on
  Service-Oriented Computing}}. Springer, \bibinfo{pages}{459--475}.
\newblock


\bibitem[Martin-Lopez et~al\mbox{.}(2021a)]%
        {martinlopez2021comparing}
\bibfield{author}{\bibinfo{person}{Alberto Martin-Lopez},
  \bibinfo{person}{Sergio Segura}, {and} \bibinfo{person}{Antonio
  Ruiz-Cort\'{e}s}.} \bibinfo{year}{2021}\natexlab{a}.
\newblock \showarticletitle{{Black-Box and White-Box Test Case Generation for
  RESTful APIs: Enemies or Allies?}}. In \bibinfo{booktitle}{\emph{Proceedings
  of the 32nd International Symposium on Software Reliability Engineering}}.
\newblock
\newblock
\shownote{to appear}.


\bibitem[Newman(2015)]%
        {newman:2015}
\bibfield{author}{\bibinfo{person}{Sam Newman}.}
  \bibinfo{year}{2015}\natexlab{}.
\newblock \bibinfo{booktitle}{\emph{Building Microservices}
  (\bibinfo{edition}{1st} ed.)}.
\newblock \bibinfo{publisher}{O’Reilly Media}.
\newblock
\showISBNx{1491950358}


\bibitem[NLTK(2021)]%
        {nltk}
NLTK \bibinfo{year}{2021}\natexlab{}.
\newblock \bibinfo{title}{NLTK}.
\newblock
\newblock
\urldef\tempurl%
\url{https://www.nltk.org/}
\showURL{%
\tempurl}
\newblock
\shownote{Accessed: Jun 3, 2022}.


\bibitem[OpenAPI(2022)]%
        {openapi}
OpenAPI \bibinfo{year}{2022}\natexlab{}.
\newblock \bibinfo{title}{OpenAPI Specification}.
\newblock
\newblock
\urldef\tempurl%
\url{https://swagger.io/specification/}
\showURL{%
\tempurl}
\newblock
\shownote{Accessed: Jun 3, 2022}.


\bibitem[Pinheiro et~al\mbox{.}(2013)]%
        {pinheiro2013model}
\bibfield{author}{\bibinfo{person}{Pedro Victor~Pontes Pinheiro},
  \bibinfo{person}{Andre~Takeshi Endo}, {and} \bibinfo{person}{Adenilso
  Simao}.} \bibinfo{year}{2013}\natexlab{}.
\newblock \showarticletitle{Model-based testing of RESTful web services using
  UML protocol state machines}. In \bibinfo{booktitle}{\emph{Brazilian Workshop
  on Systematic and Automated Software Testing}}. Citeseer,
  \bibinfo{pages}{1--10}.
\newblock


\bibitem[Postman(2022)]%
        {postman}
Postman \bibinfo{year}{2022}\natexlab{}.
\newblock \bibinfo{title}{Postman}.
\newblock
\newblock
\urldef\tempurl%
\url{https://getpostman.com}
\showURL{%
\tempurl}
\newblock
\shownote{Accessed: Jun 3, 2022}.


\bibitem[progweb(2022)]%
        {progweb}
progweb \bibinfo{year}{2022}\natexlab{}.
\newblock \bibinfo{title}{ProgrammableWeb API Directory}.
\newblock
\newblock
\urldef\tempurl%
\url{https://www.programmableweb.com/category/all/apis}
\showURL{%
\tempurl}
\newblock
\shownote{Accessed: Jun 3, 2022}.


\bibitem[raml(2022)]%
        {raml}
raml \bibinfo{year}{2022}\natexlab{}.
\newblock \bibinfo{title}{RESTful API Modeling Language}.
\newblock
\newblock
\urldef\tempurl%
\url{https://raml.org/}
\showURL{%
\tempurl}
\newblock
\shownote{Accessed: Jun 3, 2022}.


\bibitem[ReadyAPI(2022)]%
        {readyapi}
ReadyAPI \bibinfo{year}{2022}\natexlab{}.
\newblock \bibinfo{title}{ReadyAPI}.
\newblock
\newblock
\urldef\tempurl%
\url{https://smartbear.com/product/ready-api/overview/}
\showURL{%
\tempurl}
\newblock
\shownote{Accessed: Jun 3, 2022}.


\bibitem[RESTAssured(2022)]%
        {restassured}
RESTAssured \bibinfo{year}{2022}\natexlab{}.
\newblock \bibinfo{title}{REST Assured}.
\newblock
\newblock
\urldef\tempurl%
\url{https://rest-assured.io}
\showURL{%
\tempurl}
\newblock
\shownote{Accessed: Jun 3, 2022}.


\bibitem[RESTest(2022)]%
        {restesttool}
RESTest \bibinfo{year}{2022}\natexlab{}.
\newblock \bibinfo{title}{RESTest}.
\newblock
\newblock
\urldef\tempurl%
\url{https://github.com/isa-group/RESTest}
\showURL{%
\tempurl}
\newblock
\shownote{Accessed: Jun 3, 2022}.


\bibitem[RESTler(2022)]%
        {restlertool}
RESTler \bibinfo{year}{2022}\natexlab{}.
\newblock \bibinfo{title}{RESTler}.
\newblock
\newblock
\urldef\tempurl%
\url{https://github.com/microsoft/restler-fuzzer}
\showURL{%
\tempurl}
\newblock
\shownote{Accessed: Jun 3, 2022}.


\bibitem[Reza and Van~Gilst(2010)]%
        {reza2010framework}
\bibfield{author}{\bibinfo{person}{Hassan Reza} {and} \bibinfo{person}{David
  Van~Gilst}.} \bibinfo{year}{2010}\natexlab{}.
\newblock \showarticletitle{A framework for testing RESTful web services}. In
  \bibinfo{booktitle}{\emph{2010 Seventh International Conference on
  Information Technology: New Generations}}. IEEE, \bibinfo{pages}{216--221}.
\newblock


\bibitem[Saad et~al\mbox{.}(2019)]%
        {saad2019exploring}
\bibfield{author}{\bibinfo{person}{Muhammad Saad}, \bibinfo{person}{Jeffrey
  Spaulding}, \bibinfo{person}{Laurent Njilla}, \bibinfo{person}{Charles
  Kamhoua}, \bibinfo{person}{Sachin Shetty}, \bibinfo{person}{DaeHun Nyang},
  {and} \bibinfo{person}{Aziz Mohaisen}.} \bibinfo{year}{2019}\natexlab{}.
\newblock \showarticletitle{Exploring the attack surface of blockchain: A
  systematic overview}.
\newblock \bibinfo{journal}{\emph{arXiv preprint arXiv:1904.03487}}
  (\bibinfo{year}{2019}).
\newblock


\bibitem[schemathesis(2022)]%
        {schemathesis}
schemathesis \bibinfo{year}{2022}\natexlab{}.
\newblock \bibinfo{title}{schemathesis}.
\newblock
\newblock
\urldef\tempurl%
\url{https://github.com/schemathesis/schemathesis}
\showURL{%
\tempurl}
\newblock
\shownote{Accessed: Jun 1, 2022}.


\bibitem[Segura et~al\mbox{.}(2017)]%
        {segura2017metamorphic}
\bibfield{author}{\bibinfo{person}{Sergio Segura}, \bibinfo{person}{Jos{\'e}~A
  Parejo}, \bibinfo{person}{Javier Troya}, {and} \bibinfo{person}{Antonio
  Ruiz-Cort{\'e}s}.} \bibinfo{year}{2017}\natexlab{}.
\newblock \showarticletitle{Metamorphic testing of RESTful web APIs}.
\newblock \bibinfo{journal}{\emph{IEEE Transactions on Software Engineering
  (TSE)}} (\bibinfo{year}{2017}), \bibinfo{pages}{1083--1099}.
\newblock


\bibitem[Sneed and Huang(2006)]%
        {sneed2006wsdltest}
\bibfield{author}{\bibinfo{person}{Harry~M Sneed} {and}
  \bibinfo{person}{Shihong Huang}.} \bibinfo{year}{2006}\natexlab{}.
\newblock \showarticletitle{Wsdltest-a tool for testing web services}. In
  \bibinfo{booktitle}{\emph{2006 Eighth IEEE International Symposium on Web
  Site Evolution (WSE'06)}}. IEEE, \bibinfo{pages}{14--21}.
\newblock


\bibitem[Stallenberg et~al\mbox{.}(2021)]%
        {stallenberg2021improving}
\bibfield{author}{\bibinfo{person}{Dimitri Stallenberg},
  \bibinfo{person}{Mitchell Olsthoorn}, {and} \bibinfo{person}{Annibale
  Panichella}.} \bibinfo{year}{2021}\natexlab{}.
\newblock \showarticletitle{Improving Test Case Generation for REST APIs
  Through Hierarchical Clustering}. In \bibinfo{booktitle}{\emph{2021 36th
  IEEE/ACM International Conference on Automated Software Engineering (ASE)}}.
  IEEE, \bibinfo{pages}{117--128}.
\newblock


\bibitem[tcases(2022)]%
        {tcases}
tcases \bibinfo{year}{2022}\natexlab{}.
\newblock \bibinfo{title}{tcases restapi tool}.
\newblock
\newblock
\urldef\tempurl%
\url{https://github.com/Cornutum/tcases/tree/master/tcases-openapi}
\showURL{%
\tempurl}
\newblock
\shownote{Accessed: Jun 3, 2022}.


\bibitem[Tsai et~al\mbox{.}(2002)]%
        {tsai2002coyote}
\bibfield{author}{\bibinfo{person}{Wei-Tek Tsai}, \bibinfo{person}{Ray Paul},
  \bibinfo{person}{Weiwei Song}, {and} \bibinfo{person}{Zhibin Cao}.}
  \bibinfo{year}{2002}\natexlab{}.
\newblock \showarticletitle{Coyote: An xml-based framework for web services
  testing}. In \bibinfo{booktitle}{\emph{Proceedings of 7th IEEE International
  Symposium on High Assurance Systems Engineering}}. IEEE,
  \bibinfo{pages}{173--174}.
\newblock


\bibitem[Viglianisi et~al\mbox{.}(2020)]%
        {viglianisi2020resttestgen}
\bibfield{author}{\bibinfo{person}{Emanuele Viglianisi},
  \bibinfo{person}{Michael Dallago}, {and} \bibinfo{person}{Mariano Ceccato}.}
  \bibinfo{year}{2020}\natexlab{}.
\newblock \showarticletitle{RestTestGen: automated black-box testing of RESTful
  APIs}. In \bibinfo{booktitle}{\emph{2020 IEEE 13th International Conference
  on Software Testing, Validation and Verification (ICST)}}. IEEE,
  \bibinfo{pages}{142--152}.
\newblock


\bibitem[Vosta(2020)]%
        {vosta2020evaluation}
\bibfield{author}{\bibinfo{person}{Diba Vosta}.}
  \bibinfo{year}{2020}\natexlab{}.
\newblock \bibinfo{title}{Evaluation of the t-wise Approach for Testing REST
  APIs}.
\newblock
\newblock


\bibitem[Voutilainen(2003)]%
        {voutilainen2003part}
\bibfield{author}{\bibinfo{person}{Atro Voutilainen}.}
  \bibinfo{year}{2003}\natexlab{}.
\newblock \showarticletitle{Part-of-speech tagging}.
\newblock \bibinfo{journal}{\emph{The Oxford handbook of computational
  linguistics}} (\bibinfo{year}{2003}), \bibinfo{pages}{219--232}.
\newblock


\bibitem[WAS(2022)]%
        {was}
WAS \bibinfo{year}{2022}\natexlab{}.
\newblock \bibinfo{title}{Qualys Web Application Scanning (WAS)}.
\newblock
\newblock
\urldef\tempurl%
\url{https://www.qualys.com/apps/web- app- scanning/}
\showURL{%
\tempurl}
\newblock
\shownote{Accessed: Jun 3, 2022}.


\bibitem[Webster and Kit(1992)]%
        {webster1992tokenization}
\bibfield{author}{\bibinfo{person}{Jonathan~J Webster} {and}
  \bibinfo{person}{Chunyu Kit}.} \bibinfo{year}{1992}\natexlab{}.
\newblock \showarticletitle{Tokenization as the initial phase in NLP}. In
  \bibinfo{booktitle}{\emph{COLING 1992 Volume 4: The 14th International
  Conference on Computational Linguistics}}.
\newblock


\bibitem[Wood et~al\mbox{.}(2014)]%
        {wood2014ethereum}
\bibfield{author}{\bibinfo{person}{Gavin Wood} {et~al\mbox{.}}}
  \bibinfo{year}{2014}\natexlab{}.
\newblock \showarticletitle{Ethereum: A secure decentralised generalised
  transaction ledger}.
\newblock \bibinfo{journal}{\emph{Ethereum project yellow paper}}
  \bibinfo{volume}{151}, \bibinfo{number}{2014} (\bibinfo{year}{2014}),
  \bibinfo{pages}{1--32}.
\newblock


\bibitem[Wu et~al\mbox{.}(2022)]%
        {wu2022combinatorial}
\bibfield{author}{\bibinfo{person}{Huayao Wu}, \bibinfo{person}{Lixin Xu},
  \bibinfo{person}{Xintao Niu}, {and} \bibinfo{person}{Changhai Nie}.}
  \bibinfo{year}{2022}\natexlab{}.
\newblock \showarticletitle{Combinatorial Testing of RESTful APIs}. In
  \bibinfo{booktitle}{\emph{ACM/IEEE International Conference on Software
  Engineering (ICSE)}}.
\newblock


\bibitem[Zhang et~al\mbox{.}(2021)]%
        {zhang2021resource}
\bibfield{author}{\bibinfo{person}{Man Zhang}, \bibinfo{person}{Bogdan
  Marculescu}, {and} \bibinfo{person}{Andrea Arcuri}.}
  \bibinfo{year}{2021}\natexlab{}.
\newblock \showarticletitle{Resource and dependency based test case generation
  for RESTful Web services}.
\newblock \bibinfo{journal}{\emph{Empirical Software Engineering}}
  (\bibinfo{year}{2021}), \bibinfo{pages}{1--61}.
\newblock


\bibitem[Zhu and Goldberg(2009)]%
        {zhu2009introduction}
\bibfield{author}{\bibinfo{person}{Xiaojin Zhu} {and} \bibinfo{person}{Andrew~B
  Goldberg}.} \bibinfo{year}{2009}\natexlab{}.
\newblock \showarticletitle{Introduction to semi-supervised learning}.
\newblock \bibinfo{journal}{\emph{Synthesis lectures on artificial intelligence
  and machine learning}} (\bibinfo{year}{2009}), \bibinfo{pages}{1--130}.
\newblock


\end{thebibliography}

\end{document}